\documentclass[journal=jacsat,manuscript=article]{achemso}
\usepackage[version=3]{mhchem} % Formula subscripts using \ce{}
\usepackage{wasysym}
\usepackage{amsmath,amssymb,latexsym}
\usepackage{pifont}
\usepackage[dvipsnames]{xcolor}
\usepackage{color}
\usepackage{tikz}
\usepackage{csquotes}
\usetikzlibrary{shapes}
\usepackage{bm}
\usepackage{subcaption}
\usepackage{caption}
\usepackage{float}

\author{M. G. Guenza}
\email{mguenza@uoregon.edu}
\affiliation{Department of Chemistry and Biochemistry, University of Oregon, Eugene, Oregon 97403}
\alsoaffiliation{Department of Physics, University of Oregon, Eugene, Oregon 97403, USA}
\alsoaffiliation{Material Science Institute, University of Oregon, Eugene, Oregon 97403, USA}
\altaffiliation{Institute for Fundamental Science, University of Oregon, Eugene, Oregon 97403, USA}

\title{Cooperative polymer dynamics at the crossover between the unentangled and the entangled regimes: theoretical predictions of scattering and linear shear relaxation for polyethylene melts}
\affiliation{Department of Chemistry and Biochemistry,  University of Oregon, Eugene, Oregon 97403}
\date{\today}
\SectionNumbersOn 
\begin{document}
For Table of Contents use only
\begin{figure}[H]
\centering
\includegraphics[width=.8\columnwidth]{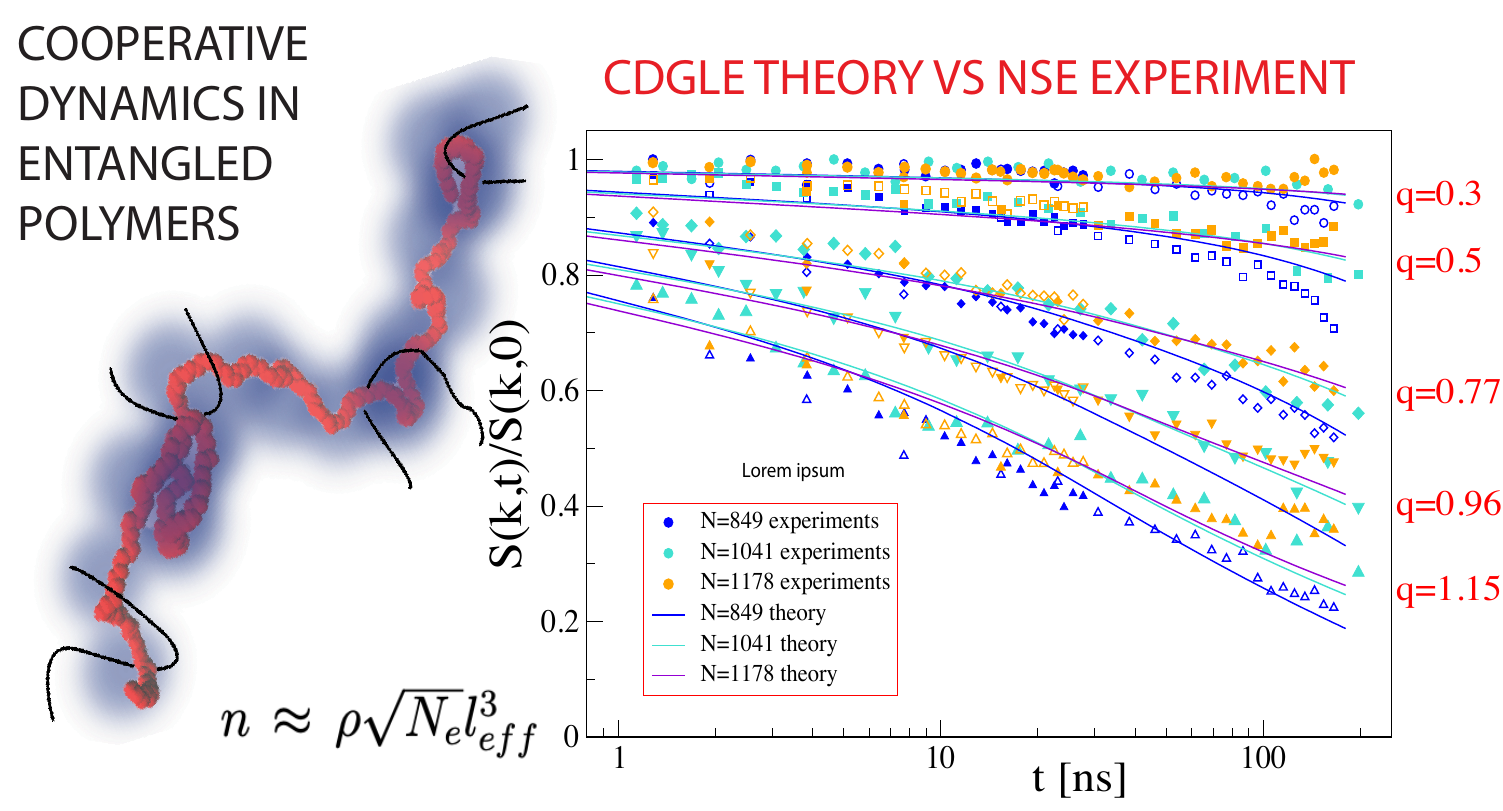}
\end{figure}

\clearpage

%\begin{tocentry}
%\includegraphics[width=1.\columnwidth]{figurepaper/TOCFINAL.pdf}
%\end{tocentry}

\begin{abstract}
%150 words
%The dynamics of polymer melts at the crossover between unentagled and entangled regimes is formalized here through an extension of the Cooperative Dynamics Generalized Langevin Equation (CDGLE) (\textit{J. Chem. Phys.} 110,7574 (1999)). The extended CDGLE incorporates the interplay between entanglements and cooperative dynamic by introducing an effective inter-monomer potential that confines the motion of the chains. The CDGLE is directly compared with Neutron Spin Echo (NSE) experiment data on polyethylene melts, demonstrating excellent agreement across both unentangled and weakly entangled regimes (up to twelve entanglements). Additionally, the approach highlights the impacts of cooperative dynamics, local flexibility, and entanglements on the CDGLE predictions of linear shear relaxation.

 The dynamics of polymer melts at the crossover between unentagled and entangled regimes is formalized here through an extension of the Cooperative Dynamics Generalized Langevin Equation (CDGLE) (\textit{J. Chem. Phys.} 110,7574 (1999)), by including the constraint to the dynamics due to entanglements through an effective inter-monomer potential that confines the motion of the chains. As one polymer chain in a melt interpenetrates with a $\sqrt{N}$ other chains, with $N$ the degree of chain polymerization, their dynamics is coupled through their potential of mean-force, leading to chains' cooperative motion and center-of-mass subdiffusive dynamics. When increasing the degree of polymerization, the extended CDGLE approach describes the dynamical behavior of unentangled to  weakly entangled systems undergoing cooperative dynamics. By direct comparison of the CDGLE with data of Neutron Spin Echo (NSE) experiments on polyethylene melts we find that the cooperative dynamics in entangled systems are confined in the region delimited by entanglements.  We extend the CDGLE to describe linear dynamical mechanical measurements and use it to calculate shear relaxation for the polyethylene samples investigated by NSE. The effects of cooperative dynamics, local flexibility, and entanglements in the shear relaxation are discussed. It is noteworthy that the theoretical approach describes with accuracy the crossover from unentangled to entangled-global dynamics for polyethylene melts of increasing chain length, covering  the regimes of unentangled and weakly entangled (up to $13$ entanglements) dynamics in one approach. 
\end{abstract}

\maketitle

\section{Introduction}
Entangled polymer melts display unique dynamical properties as the time scales of diffusion and viscosity change several orders of magnitude with increasing degree of polymerization. More specifically, the scaling exponents that define the dynamics, diffusion and viscosity, as a function of the degree of polymerization, $N$, differ notably between short and long chains. In the case of long chains, dynamics are dominated by entanglements, which are transient points of contact between chains that cause topological constraints, resulting from the chains' inability to pass through one another.\cite{Doi1988,DeGennes1971} These entanglements constrain chain diffusion, particularly when chains are very long, causing diffusion to primarily occur along the chain's curvature, resembling the slithering motion of a snake, known as "reptation".\cite{Wang2010} Conversely, in liquids comprising short chains, entanglements per chain are minimal, facilitating rapid chain diffusion, with short-lived entanglements exerting negligible impact on chain motion.\cite{Doi1988,DeGennes1971}  

Traditionally, short and long chain dynamics have been represented by two distinct and formally incompatible formalisms: the Rouse and reptation models, respectively. 

The Rouse model effectively describes the dynamics of \textit{short polymer chains} in a liquid, employing a Langevin equation to model the temporal evolution of the spatial coordinates of the monomers within a macromolecular chain.\cite{Rouse2004} The model incorporates the influence of the surrounding molecules through an effective friction coefficient and random forces. In the Rouse model, random forces are treated as uncorrelated and approximated as white noise, representing random collisions among monomers from surrounding polymer chains. This approximated description is often refined by introducing time-dependent friction via memory functions.\cite{Berne1990,Lyubimov2010, Lyubimov2011}  The Rouse model, however, does not account for the constraints due to the presence of entanglements as it represents the polymers as idealized chains that are free of cross other chains and themselves, while maintaining correct average structure, in the form of end-to-end distance and radius-of-gyration. Notably, incorporating  local semi-flexibility and cooperative dynamics enhances its agreement with experimental observations.\cite{Copperman2015,Copperman2017, Beyerle2019, Beyerle2021,Guenza1999,Guenza2002,Zamponi2008}

The dynamics of \textit{long polymer chains} in a melt is conventionally described through the reptation model, initially proposed by de Gennes and subsequently formalized by Doi and Edwards.\cite{DeGennes1971,Doi1988} In this model, fluid relaxation occurs through the anisotropic diffusion of the chain within an effective tube formed by the entangled chains surrounding the polymer, eventually leading to its final creeping outside the tube.\cite{DeGennes1971,Doi1988} The original reptation model aptly captures dynamics in the fully-entangled regime, characterized by linear chains with high polymerization degree and long-lasting entanglement constraints. However, to enhance its applicability to systems featuring shorter and weakly entangled chains, where defining a persisting confining tube becomes challenging,\cite{Ma2021} some modifications were proposed to the original reptation model. These consider additional relaxation mechanisms that encompass contour length fluctuations, constraint release, and tube dilation, acknowledging that the tube is formed by the surrounding chains moving on the same timescale than the tagged chain inside the tube.\cite{McLeish2002,LodgeRosteinPrager}  These refined reptation approaches significantly enhance agreement with experimental observations. However, they also render direct comparison with experiments challenging, as different approximations need to be applied to compare the model to different experiments, such as shear and scattering experiments. This point was clearly argued by Likhtman, who proposed the slip-link model, which mimics reptation through the numerical solution of a set of stochastic differential equations that include a phenomenological anisotropic harmonic potential.\cite{Likhtman2005} 

Despite its advancements, over the years critiques of the reptation model have emerged from both theoretical and experimental perspective. Ngai recently presented a comprehensive overview of the disparities observed between experiments and the reptation model and provided an alternative explanation of the observed phenomena using the Coupling Model.\cite{Ngai2024,Ngai2024b} 

Computer simulations of polymer melts suggest that the overall dynamics of polymers are even more complex than previously thought. They have revealed how polymer dynamics is heterogeneous, with chains forming interconverting regions of slow and fast dynamics.\cite{Guenza2022} Building on this observation, we derived a Generalized Langevin Equation for Cooperative Dynamics (CDGLE) from the Hamiltonian of the liquid by projecting the dynamics onto the coordinates of the subensemble of  slow and interacting macromolecules, moving in the field of the faster surrounding polymers.\cite{Guenza1999,Guenza2002} The resulting Generalized Langevin Equation adequately explains the center-of-mass sub-diffusive motion observed in unentangled polymer melts through neutron spin echo (NSE) experiments.\cite{Zamponi2008} In the CDGLE intramolecular and intermolecular forces govern polymer dynamics until interdiffusion leads to a loss of dynamical correlation. More details on this approach can be found in \textcolor{blue}{Section\ref{backgrounddynamics}}.

In this manuscript we expand upon the original CDGLE approach to describe the dynamics of polymer melts across varying chain lengths, including the crossover to entangled dynamics. We derive an effective anharmonic potential that acts between monomers of interacting chains within de Gennes' correlation hole.\cite{DeGennes1971} The potential is isotropic and applies to unentangled and weakly entangled systems. Preliminary results from this model were outlined in a previous short publication;\cite{Guenza2014} In this manuscript, we present a theoretical framework for entangled dynamics, including the derivation of the anharmonic, isotropic effective potential that confines polymer dynamics due to entanglements, along with new findings on scattering and mean-square displacements.

To test the CDGLE, we select a set of Neutron Spin Echo (NSE) experiments of dynamic structure factors of polyethylene melts (also reported in the literature as PEB-2 samples) collected by Richter and coworkers.\cite{Schleger1998,Zamponi2008} These data provide detailed information on the dynamics of polymer melts in a wide range of chain lengths, comprehensive of the transition from the unentangled to the entangled dynamics up to twelve entanglements per chain. When compared against the NSE data the CDGLE demonstrates remarkable accuracy.\cite{Zamponi2008,Guenza2014} Although comparisons with NSE data have been made previously, the approach in this paper enables us to separately identify the distinct contributions to scattering relaxation from chain semiflexibility, dynamical cooperativity, and entanglements.

Finally, to evaluate the effect of dynamical cooperativity in the stress relaxation,\cite{Tung2024} we extend here the CDGLE formalism to describe dynamical mechanical measurements in the linear shear perturbation. For these calculations we utilize the same parameters we obtained from reproducing the NSE experiments to formally connect scattering and linear shear experiments.

Previous efforts have been made to describe the dynamics of a single polymer chain within a  Generalized Langevin formalism where, however, entanglements are accounted for by the integral of the memory function.\cite{Schweizer1997,Fatkullin1994} Because solving the integral of the memory function is not trivial and may yield different results depending on the type of approximations used, the CDGLE abandons the single-chain perspective and instead models the effect of entanglements as an interchain potential between monomers in a group of 
$n$ interacting chains.\cite{Guenza1998,Guenza1999}

The topological constraints arising from entanglements are formalized starting from the intermolecular monomer pair distribution function, leading to a time-dependent potential.  In our model, an ensemble of $n$ interacting polymers exhibits cooperative motion within the volume defined by the correlation hole. The entangled chains within this ensemble experience a confining potential that evolves over time as the molecules interdiffuse.\cite{Robertson2007,Graham2003}

Although our entanglement potential is harmonic at any given time, the time-dependent nature of the related force constant results in an effective time-dependent potential that is anharmonic, consistently with both simulations\cite{Zhou2006,Ramanathan} and experiments.\cite{Robertson2007,Wang2010,Glaser2010} It is important to note that an \textit{exact} microscopic formalism for the potential confining highly entangled systems is only feasible for rods or polymers represented as chains of coarse-grained stiff needles, as derived by Sussman and Schweizer.\cite{Sussman2011,Sussman2012} In their approach, the potential is anisotropic and effectively captures the dynamics of strongly entangled systems where the motion is directionally dependent. Since we do not assume any anisotropy in the formalism,  our theory remains isotropic and applies only to unentangled and weakly-entangled systems where the dynamics are isotropic.   

The same CDGLE approach applies to unentangled and weakly entangled systems.
Entangled dynamics naturally emerge in the theoretical predictions of CDGLE for samples of long polymer chains where interdiffusion is slow, while the entanglement potential does not impact the dynamics of short chains that interdiffuse rapidly. Additionally, the CDGLE preserves the formal structure of the Rouse equation and its mathematically convenient representation in normal modes.\cite{Guenza1999,Guenza2002}

Among the most intriguing predictions of the CDGLE theory are as follows: i) the existence of cooperative motions in polymer melts, leading to subdiffusive dynamics, irrespective of the monomeric structure and local semiflexibility \cite{Guenza2022}. The CDGLE presents a comprehensive framework with the polymer persistence length as a key parameter, thus encompassing systems with diverse monomeric structures; and ii) the ubiquity of cooperative dynamics across all polymer melts regardless of their degree of polymerization. 
The relevance of cooperative motion emerging from intermolecular correlation has been highlighted in recent work by Wang and coworkers through a detailed analysis of coarse-grained MD simulations and scattering experiments, both in the real and reciprocal space.\cite{Ma2021,Shen2022} The presence of cooperative motion has been hypothesized to explain the experimentally-observed subdiffusive dynamics of intrinsically disordered proteins within coacervates.\cite{Galvanetto2023,Guenza2023}
Finally, comparison with NSE experiments reveals that iii) in entangled systems, the confinement imposed by entanglements constraints the region of cooperativity to the volume delineated by these entanglement constraints.\cite{Guenza2014} 

Several of these predictions closely align with recent findings from Neutron Spin Echo (NSE) studies, where cooperative dynamics of unentangled and entangled chains were thoroughly investigated by Richter, Kruteva, Zamponi, and their collaborators \cite{Richter2024, Monkenbusch2023, Sharma2022a, Richter2021a}. Considering that the CDGLE has been applied for comparison and testing to their NSE experiments, this alignment comes as no surprise.

The paper is structured as follows: \textcolor{blue}{Section \ref{backgrounddynamics}} provides a summary of the Langevin equation governing the cooperative dynamics in both unentangled and entangled polymer melts, where \textcolor{blue}{Subsection \ref{timecorrelationfunctions}} formally derives the time correlation functions needed to compare the theory with experiments. This is followed by \textcolor{blue}{Section \ref{potential}}, which presents the derivation of the effective potentials incorporated into the CDGLE, including the confining potential arising from entanglements.  The numerical self-consistent procedure utilized to solve the CDGLE equation, along with the method for determining its parameters and their values, are outlined in \textcolor{blue}{Section \ref{numerical}}. \textcolor{blue}{Section \ref{NSE}} compares the CDGLE against Neutron Spin Echo data, encompassing an analysis of various contributions to the dynamic structure factor, such as semiflexibility, cooperativity, non-zero $\alpha$ parameter, and entanglements. \textcolor{blue}{Section \ref{shear}} introduces the CDGLE theory of stress relaxation under linear shear and predicts stress relaxation for the samples studied via NSE. \textcolor{blue}{Section \ref{AppendixA}} reports the analytical solution of the entanglement force, while \textcolor{blue}{Section \ref{AppendixB}} presents the derivation of the shear relaxation modulus under a linear perturbation for the CDGLE. The paper summarizes our findings in the concluding section.

\section{The Many-Chain Cooperative Dynamics model from unentangled to entangled dynamics}
\label{backgrounddynamics}
This paper builds upon the Cooperative Dynamics Generalized Langevin Equation (CDGLE) theory for polymeric liquids \cite{Guenza1999, Guenza2002}, extending the original framework to describe cooperative many-chain dynamics across both unentangled and entangled regimes. In this section we briefly present the essential equations that are needed to calculate mean-square displacements, dynamic structure factors, and shear relaxation. The extended theoretical details of the new developments are provided in the \textit{Supporting Information}, while the original theory for unentangled chains is illustrated in our previous publications.\cite{Guenza1998, Guenza2002}

The fundamental concept in the CDGLE is that in a polymer liquid characterized by a monomer density $\rho$ and chains with degree of polymerization $N$, approximately $n\propto \rho \sqrt{N}$ chains are enclosed within a volume defined by the range of the potential of mean force, which corresponds to de Gennes' correlation hole.\cite{DeGennes1971} The radius of this spherical volume is proportional to the radius of gyration $R_g\propto  N^{1/2}$ of any given chain. Pair of chains in the $n$ ensemble are interacting at the center-of-mass level through the potential of mean force, leading to the center-of-mass subdiffusive motion.\cite{Guenza1999,Guenza2002}

Molecular dynamics (MD) simulations of polyethylene melts demonstrate that these $n$ interpenetrating chains gradually lose correlation over time as they interdiffuse.\cite{Guenza2022} This is evidenced by their van Hove time correlation function decaying to unity after a characteristic time, $\tau_{corr}$, which represents the time required for the chains to diffuse a distance comparable to the range of the mean-force potential, and roughly corresponds to the longest Rouse relaxation time.

In the theory presented here, at the monomer level, correlated chains interact through an interchain potential that mimics the confinement of entanglements. Notably, the intermolecular center-of-mass and monomer forces are not inserted through an \textit{ad hoc} procedure in the CDGLE equation; rather, they emerge from the Mori-Zwanzig projection operator method as van Hove functions, and are subsequently approximated.\cite{Zwanzig1974,Guenza1999} The interactions are, at any given time, harmonic, but the corresponding spring constants are time-dependent, yielding effective anharmonic pair potentials that guide the dynamics.

At each time interval, these constants are optimized through a self-consistent procedure until convergence of the interpolymer and intermonomer distances is achieved, leading to effective anharmonic potentials (see \textcolor{blue}{Section\ref{potential}}). 

 Given a group of $n$ interpenetrating polymers, which initially occupy the volume of the correlation hole, the time evolution of a generic monomer $i$ in the polymer $a$, which is interacting with polymer $b$ belonging to another chain, is governed by a Langevin equation in the space coordinates
\begin{eqnarray}
\label{eq:langevin}
\zeta_{eff} \frac{d \mathbf{r}_i^{(a)}(t)}{d t} & = & \frac{3}{ \beta l^2} 
\sum_j^N \mathbf{A}_{i,j} \mathbf{r}_j^{(a)}(t)-(n-1)K[r(t)] \mathbf{r}_i^{(a)}(t) 
+  \sum_{b \neq a}^{n} K[r(t)] \mathbf{r}_{cm}^{(b)}(t)
\label{eq:langevinnomemory} 
+ \mathbf{F}_i^{Q(a)}(t)  \ .
\end{eqnarray}
In the spirit of the Rouse model, the equation contains viscous forces, $\zeta_{eff} \frac{d \mathbf{r}_i^{(a)}(t)}{d t}$, intramolecular forces, $\frac{3}{ \beta l^2} 
\sum_j^N \mathbf{A}_{i,j} \mathbf{r}_j^{(a)}(t)$, random forces, $\mathbf{F}_i^{Q(a)}(t)$, but with the inclusion of time-dependent intermolecular forces, with
$K[r(t)]$ the force constants. The methodology for calculating these intermolecular forces is detailed in \textcolor{blue}{Section \ref{potential}}.

Taking advantage of the statistical equivalence of each chain in the sub-ensemble and the isotropic nature of the liquid, the solution of the coupled equations reduces to studying the dynamics of a pair of chains. More specifically, Eq.\ref{eq:langevin} is solved using a similarity transformation, which separates the formalism into $(n$ $-$ $1)$ identical equations in the relative monomer coordinates, $\mathbf{r}_D(t)= (\mathbf{r}^{(a)}(t)-\mathbf{r}^{(b)}(t))/\sqrt{2}$, and one equation in the collective monomer coordinates, $\mathbf{r}_N(t)=[\sum_{i=1}^n\mathbf{r}^{(i)}(t)]/\sqrt{n}$ as
\begin{eqnarray}
\zeta_{eff} \frac{d \mathbf{r}^D(t)}{dt}= - k_s \mathbf{A}^D \mathbf{r}^D(t) + \mathbf{F}_D^{Q}(t) \  \\
\zeta_{eff} \frac{d \mathbf{r}^N(t)}{dt}= - k_s \mathbf{A}^N \mathbf{r}^N(t) + \mathbf{F}_N^{Q}(t) \ ,
\end{eqnarray}
with
\begin{eqnarray}
\label{AD}
\mathbf{A}^D=\frac{\beta l^2 K[r(t)]}{3} [ (n-1)\mathbf{1}+\mathbf{Q}_0 \mathbf{Q}_0^T  ] + \mathbf{A} \ ,
\end{eqnarray}
\begin{eqnarray}
\label{AN}
\mathbf{A}^N=\frac{(n-1)\beta l^2 K[r(t)]}{3} [ \mathbf{1}-\mathbf{Q}_0\mathbf{Q}_0^T ] + \mathbf{A} \ ,
\end{eqnarray}
where $\mathbf{A}$ is the single chain intramolecular matrix, corresponding to the Rouse matrix for a fully flexible polymer. Nevertheless, because real polymers must be modeled with more realistic intramolecular forces than those described by the Rouse model, CDGLE uses a matrix formalism to describe semiflexible chains of finite length represented as freely rotating chains (FRC).\cite{Guenza1999} %If the polymer radius-of-gyration is  $R_g^2=Nl_{eff}^2/6$, the effective segment is  $\mathbf{l}_{eff} \approx \mathbf{l} \ (1 + g)/(1 - g)$, where  $g=- \langle\cos \theta\rangle$ and $l=1.53 \AA$ represents the carbon-carbon bond length. The parameter $\theta$ is the statistical angle between two consecutive carbon-carbon bonds, accounting for the polymer's local semiflexibility \cite{Ramachandran2008}. This formalism is applicable to polymeric chains with variable monomeric structures, with the $\theta$ parameter depending on the specific polymer under study.\cite{Flory1970,Lyubimov2011,Lyubimov2010} For the polyethylene melt studied here, $g = 0.785$. This model reverts to the Rouse intramolecular force when $g = 0$.\cite{Doi1988} 

The eigenvector $\mathbf{Q}_0$ is the first eigenvector of the Rouse matrix, defined as $\mathbf{Q}_0^T=N^{-1/2}(1, 1, ...., 1)$. By adopting an identical friction coefficient, $\zeta_{eff}$, for both relative and collective coordinates we assume that the relative and collective memory functions in the original Generalized Langevin Equation are Markovian and integrated to give the effective friction coefficient.\cite{Zwanzig1974,Guenza1998} This approximation is valid when the most relevant slow dynamics are already properly accounted for in the linearized part of the equation, as we argue is the case here.\cite{Zwanzig1974}

\subsection{Solution of the CDGLE in normal modes}
\label{modes}

The CDGLE formalism can be conveniently solved  by transformation into independent normal modes, consistently with the Rouse model. This semplification is possible because the intramolecular and the intermolecular potentials, including the confinement due to entanglements, are modeled as harmonic springs, with the caveat that the intermolecular spring constants are time dependent.  Entanglements confine the relative chain motion at the monomer level, while cooperative dynamics mainly affect the center-of-mass interdiffusion.

For the \textit{center of mass dynamics}, where $p=0$ and $\lambda_0=0$, the motion in relative and collective coordinates follows the equations
\begin{eqnarray}
\label{eq:oho}
\zeta_{eff} \frac{d \mathbf{\xi}_{0}(t)}{d t} & = & - k_s \Lambda^D_0\mathbf{\xi}_{0}(t)
+ \mathbf{F}_{0}^{\xi}(t) \ ,  \\
\zeta_{eff} \frac{d \mathbf{\chi}_{0}(t)}{d t} & = & \mathbf{F}_{0}^{\chi}(t)  \ , \nonumber
\end{eqnarray}
with 
\begin{eqnarray}
\label{lambda0}
    \Lambda^D_0=n K_0[r(t)]/k_s \ ,
\end{eqnarray} 
and $n K_0[r(t)]$ the repulsive force constant responsible for correlated chain dynamics, defined in \textcolor{blue}{Section \ref{potential}}. The collective dynamics of the group of chains is diffusive, $\Lambda^N_0=0$, and the fluctuation-dissipation theorems are defined as\cite{Guenza1999}
\begin{eqnarray}
\label{perrocche'}
\langle \mathbf{F}_{0}^{\xi}(t) \cdot \mathbf{F}_{0}^{\xi}(t')\rangle & = & 6 n(n-1) k_B T \zeta_{eff} \delta(t,t')  \ , \label{lillibet} \\
\langle \mathbf{F}_{0}^{\chi}(t) \cdot \mathbf{F}_{0}^{\chi}(t')\rangle & = & 6 n  \ k_B T \zeta_{eff} \delta(t,t')  \ . \nonumber
\end{eqnarray}
Eqs.(\ref{perrocche'}) emerge when the multibody structural distribution function is approximated by a product of pair distribution functions. Thus, the single-chain center-of-mass dynamics  is subdiffusive until the polymer moves a relative distance, $r(t)$, larger than the range of the interaction potential, when the intermolecular interaction at the center-of-mass level reduces to zero, $K_0[r(t)]=0$, and Eq.(\ref{eq:oho}) reduces to the  well-known diffusion equation. 
%For short, unentangled chains this limit corresponds to the  trivial motion of $n$ independent, inter-diffusing chains, each following the Rouse equation with internal semiflexibility.

 The \textit{internal dynamics} ($p=1,\ 2, ...,N-1$)  are expressed as a set of uncoupled equations of motion in the relative, $\mathbf{\xi}_{p}^{(a)}(t)$, and collective, $\mathbf{\chi}_{p}^{(a)}(t)$, normal mode coordinates,
\begin{eqnarray}
\label{eq:normalm11}
\zeta_{eff} \frac{d \mathbf{\xi}_{p}^{(a)}(t)}{d t} & = & - k_s \Lambda^D_p  \mathbf{\xi}_{p}^{(a)}(t)
 +  \mathbf{F}_{p}^{\xi}(t)  \ , \\
\zeta_{eff} \frac{d \mathbf{\chi}_{p}^{(a)}(t)}{d t} & = & - k_s \Lambda^N_p \mathbf{\chi}_{p}^{(a)}(t)
 +  \mathbf{F}_{p}^{\chi}(t)  \ ,
\end{eqnarray}
with 
\begin{equation}
\label{lambdadn}
\Lambda^D_p=\Lambda^N_p=\lambda_{p} +(n-1)K[r(t)]/k_s
\end{equation}
where $\lambda_{p}$ are the eigenvalues of the single-chain intramolecular matrix: if the chain is fully flexible, $\lambda_{p}$ are the Rouse eigenvalues. $K[r(t)]$ is the time-dependent attractive force constant, defined in \textcolor{blue}{Section \ref{potential}},  due to entanglements acting between monomers belonging to different entangled chains inside the ensemble of $n$ interpenetrating polymers. Here, the relative mode coordinates are defined as 
$\mathbf{\xi}_{p}(t)=\sum_i \left[ \mathbf{Q}_{i,p} \right]^{-1} \mathbf{r}^D_{i}(t)$, while the collective mode coordinates are $\mathbf{\chi}_{p}(t)=\sum_i \left[ \mathbf{Q}_{i,p} \right]^{-1} \mathbf{r}^N_{i}(t)$.  

\noindent The related fluctuation-dissipation theorem reads
\begin{eqnarray}
\label{fluctdiss}
\langle \mathbf{F}_{p}(t) \cdot \mathbf{F}_{q}(t')\rangle = 6 k_B T \zeta_{eff} \delta(t,t') \delta(p,q)  \ ,
\end{eqnarray}
for both the relative ($\xi_p$) and the collective ($\chi_p$) dynamics.

The solution of Eqs. (\ref{eq:normalm11}) for the internal modes ($p=1, \ 2, ..., N$)  leads to the equations in the relative and collective coordinates, respectively,
\begin{eqnarray}
\label{eq:coordinates0}
\mathbf{\xi}_p (t) & = & \mathbf{\xi}_p(0)e^{- t/\tau_{\xi,p}(t)} 
 +  e^{- t/\tau_{\xi,p}(t)}  \int_{0}^{t} d \tau \zeta_{eff}^{-1} \mathbf{F}^{\xi}_{p}(\tau) e^{\tau/\tau_{\xi,p}(\tau)} \ , \\
\mathbf{\chi}_p (t) & = & \mathbf{\chi}_p(0)e^{- t/\tau_{\chi,p}(t)} 
 +  e^{- t/\tau_{\chi,p}(t)}  \int_{0}^{t} d \tau \zeta_{eff}^{-1} \mathbf{F}^{\chi}_{p}(\tau) e^{\tau/\tau_{\chi,p}(\tau)} \ , 
\end{eqnarray}

with 
\begin{eqnarray}
\label{tdsc}
\frac{t}{\tau_{\xi,p}(t)}=\frac{t}{\tau_{\chi,p}(t)}=\frac{k_s \lambda_{p}t}{\zeta_{eff}}+\frac{(n-1)}{\zeta_{eff}}\int_0^t K[r(\tau)]d\tau \  ,
\end{eqnarray}

\noindent where $\tau_{single-chain,p}=\zeta_{eff}/(k_s \lambda_{p})$ is the characteristic relaxation time of the single chain without entanglement effects.

If the chain is fully relaxed before this characteristic time at which chains start to feel the presence of entanglements, i.e. if $\tau_{decorr} < \tau_e$ with $\tau_{decorr}=\tau_{single-chain,p=1}$ the relaxation time of the longest single-chain mode, and  $\tau_e=d^2/D$ the entanglement time, the entanglements are not affecting the dynamics. This is the case for unentangled chains in our model, while for entangled chains where $\tau_{decorr} > \tau_e$ the chains feel the effect of the entanglement potential.

By applying the eigenvector transformation and the inverse of the similarity transformation matrix (see \textcolor{blue}{SI}), the  dynamics in \textit{real space} coordinates is defined as a function of the normal modes through the transformation
\begin{eqnarray}
\label{monomerposition}
\mathbf{r}^a_i(t)  =  \sum_{p=0}^{n(N-1)}  \mathbf{Q'}_{i,p} \mathbf{x}^a _p(t)  =  \sum_{k=1}^n \frac{1}{\sqrt{k(k+1)}}  \sum_{p=0}^{N-1}  (\mathbf{Q}_\xi)_{i,p} [\mathbf{\xi}_p(t)] +  \frac{1}{\sqrt{n}}\sum_{p=0}^{N-1}  (\mathbf{Q}_\chi)_{i,p} [\mathbf{\chi}_p(t)] \ , 
\end{eqnarray}
where $(\mathbf{Q}_\xi)_{i,p}$ and $(\mathbf{Q}_\chi)_{i,p}$ are the eigenvectors of the relative and collective equations of motion, respectively.

\subsection{Time correlation functions in the CDGLE approach}
\label{timecorrelationfunctions}
This section presents the essential time correlation functions (TCFs) required for calculating the monomer and center-of-mass mean-square displacements, the dynamic structure factor, and the shear modulus.

The center-of-mass mean-square displacement (\textcolor{blue}{Section \ref{msds}}) 
\begin{eqnarray}
\label{drcm}
\Delta R^2(t)=\langle[\mathbf{r}_{cm}(t)-\mathbf{r}_{cm}(0)]^2\rangle  =  n^{-2}N^{-1} [\langle[\mathbf{\xi}_0 (t)-\mathbf{\xi}_0 (0)]^2\rangle  + \langle[\mathbf{\chi}_0 (t)-\mathbf{\chi}_0 (t)]^2\rangle] \ , 
\end{eqnarray}
where
\begin{eqnarray}
\langle[\mathbf{\xi}_0 (t) - \mathbf{\xi}_0 (0)]^2\rangle   & = &  \langle\mathbf{\xi}_0(0)^2\rangle  [e^{-  t/\tau_{\xi,0}(t)}-1]^2  + \frac{6 n(n-1) k_B T}{\zeta_{eff}}   e^{- 2 t/\tau_{\xi,0}(t)}  \int_{0}^{t} d \tau  e^{2 \tau/\tau_{\xi,0}(\tau)} \ \nonumber , \\
\langle[\mathbf{\chi}_0 (t)-\mathbf{\chi}_0 (0)]^2\rangle & = & \frac{6 n k_B T}{\zeta_{eff}} t \ , 
\end{eqnarray}
with  $\langle\mathbf{\xi}_0 (0)^2\rangle =6 n (n-1) k_BT/\zeta_{eff}$ and $\langle\mathbf{\chi}_0 (0)^2\rangle =6 n k_BT/\zeta_{eff}$. 
In the limit of non-interacting chains and at long time, Eq. \ref{drcm} correctly recovers Brownian diffusion with $\Delta R^2(t)= \frac{6 k_B T}{\zeta_{eff}} t$. 

The monomer mean-square displacement (\textcolor{blue}{Section \ref{msds}})  is
\begin{eqnarray}
\label{eqmonomermsd}
 \langle[\mathbf{r}^a_i(t) -\mathbf{r}^a_i(0)]^2\rangle  =  \frac{n-1}{n}  \sum_{p=0}^{N-1} \mathbf{Q}^2_{i,p} \langle[\mathbf{\xi}_p(t)-\mathbf{\xi}_p(0)]^2\rangle   + \frac{1}{n}  \sum_{p=0}^{N-1} \mathbf{Q}^2_{i,p}  \langle[\mathbf{\chi}_p(t)-\mathbf{\chi}_p(0)]^2\rangle  \ , 
\end{eqnarray}
where we enforced the property that the eigenvectors are orthonormal and the identity $\sum_{k=0}^{n-1} [k (k+1)]^{-1}=(n-1)/n$.
The solution of Eq.\ref{eqmonomermsd} requires  \textit{internal modes} TCFS, where $p=1, 2, ..., N-1$. For
the relative displacement of two monomers inside a pair of polymers during a time interval $\Delta t=t-t_0$, with $t_0=0$, the relative and collective displacements   
\begin{eqnarray}
\langle[\mathbf{\xi}_p (t)-\mathbf{\xi}_p (0)]^2\rangle = \langle\mathbf{\xi}_p(0)^2\rangle [e^{- t/\tau_{\xi,p}(t)} -1 ]^2  
 +  \frac{6 k_B T}{\zeta_{eff}} e^{- 2 t/\tau_{\xi,p}(t)}  \int_{0}^{t} d \tau  e^{2 \tau/\tau_{\xi,p}(\tau)} \ ,
\end{eqnarray}
and
\begin{eqnarray}
\langle[\mathbf{\chi}_p (t)- \mathbf{\chi}_p (0)]^2\rangle = \langle\mathbf{\chi}_p(0)^2\rangle [e^{- t/\tau_{\chi,p}(t)} -1]^2 + \frac{6 k_B T}{\zeta_{eff}} e^{- 2 t/\tau_{\chi,p}(t)}  \int_{0}^{t} d \tau e^{2 \tau/\tau_{\chi,p}(\tau)} \ .
\end{eqnarray}
If there are not interactions between chains, or between monomers belonging to different chains, one obtains the single-chain limit by applying the condition that $n=1$, thus recovering the Rouse formalism. 

The \textit{dynamic structure factor} (\textcolor{blue}{Section \ref{SKTt}}) depends on the correlation of the displacement of two monomers belonging to the same chain
\begin{eqnarray}
\label{perskt}
\langle[\mathbf{r}^a_i(t) -\mathbf{r}^a_j(0)]^2\rangle & = &\frac{n-1}{n} \sum_{p=0}^{N-1}  \langle[\mathbf{Q}_{i,p}\mathbf{\xi}_p(t)-\mathbf{Q}_{j,p}\mathbf{\xi}_p(0)]^2\rangle + \\  && \frac{1}{n}  \sum_{p=0}^{N-1} \langle[\mathbf{Q}_{i,p}\mathbf{\chi}_p(t)-\mathbf{Q}_{j,p}\mathbf{\chi}_p(0)]^2\rangle \ . \nonumber
\end{eqnarray}
To solve these equations, one needs the TCFs for the center-of-mass motion presented above, and the products of the first normal mode coordinates at the same time, $p=0$,
\begin{eqnarray}
 \langle\mathbf{\xi}_0 (t)^2\rangle  & = & \langle\mathbf{\xi}_0(0)^2\rangle e^{-  2t/\tau_{\xi,0}(t)}  +  \frac{6 n(n-1) k_B T}{\zeta_{eff}}  e^{- 2 t/\tau_{\xi,0}(t)}  \int_{0}^{t} d \tau  e^{2 \tau/\tau_{\xi,0}(\tau)} \ \nonumber , \\
\langle\mathbf{\chi}_0 (t)^2\rangle & = & \langle\mathbf{\chi}_0 (0)^2\rangle+  \frac{6 n k_B T}{\zeta_{eff}} t \ . 
\end{eqnarray}
For the internal modes TCFs entering Eq.\ref{perskt}, where $p=1, 2, ..., N-1$, the products of the normal mode coordinates at different times are 
\begin{eqnarray}
\langle\mathbf{\xi}_p (t) \cdot \mathbf{\xi}_p (0)\rangle & = & \langle\mathbf{\xi}_p(0)^2\rangle e^{- t/\tau_{\xi,p}(t)} \ , \nonumber\\
\langle\mathbf{\chi}_p (t) \cdot  \mathbf{\chi}_p (0)\rangle & = & \langle\mathbf{\chi}_p(0)^2\rangle e^{- t/\tau_{\chi,p}(t)} \ ,  \\
\langle\mathbf{\xi}_p (t) \cdot \mathbf{\chi}_p (0)\rangle & = & 0 \nonumber \ ,
\end{eqnarray}
and at the same-time 
\begin{eqnarray}
\langle\mathbf{\xi}_p (t)^2\rangle & = & \langle\mathbf{\xi}_p(0)^2\rangle e^{- 2 t/\tau_{\xi,p}(t)} 
+ \frac{6 k_B T}{\zeta_{eff}} e^{- 2 t/\tau_{\xi,p}(t)}  \int_{0}^{t} d \tau  e^{2 \tau/\tau_{\xi,p}(\tau)} \ , \\
\langle\mathbf{\chi}_p (t)^2\rangle & = & \langle\mathbf{\chi}_p(0)^2\rangle e^{- 2 t/\tau_{\chi,p}(t)} \nonumber 
 + \frac{6 k_B T}{\zeta_{eff}} e^{- 2 t/\tau_{\chi,p}(t)}  \int_{0}^{t} d \tau e^{2 \tau/\tau_{\chi,p}(\tau)} \ . 
\end{eqnarray}
The TCFs just presented are needed to solve the shear relaxation in \textcolor{blue}{Section \ref{shear}}.

Finally, the numerical solution of the CDGLE reported in \textcolor{blue}{Section \ref{potential}} requires the evolving time-dependent distance between the center-of-mass of two interacting polymers, 
$R_{cm}^2(t) =\langle[\xi_0(t)-\xi_0(0)]^2\rangle/N$, and the time-dependent distance between two monomers belonging to two different chains, which enters the confinement potential due to entanglements, defined as
\begin{eqnarray}
\label{mondist}
\langle(\mathbf{r}^b_j(t)-\mathbf{r}^a_i(t))^2\rangle \approx \langle(\mathbf{r}^b_j(0)-\mathbf{r}^a_i(0))^2\rangle  +  \langle[(\mathbf{r}^a_i(t) -\mathbf{r}^a_i(0))  -  (\mathbf{r}^b_j(t) -\mathbf{r}^b_j(0))]^2\rangle  
\end{eqnarray}
given that we approximate  $|\mathbf{r}^a_i(t)- \mathbf{r}^b_j(t)| - |\mathbf{r}^a_i(0) - \mathbf{r}^b_j(0)| \approx  \langle[(\mathbf{r}^a_i(t) -\mathbf{r}^a_i(0)) -  (\mathbf{r}^b_j(t) -\mathbf{r}^b_j(0))]^2\rangle ^{1/2}$. Both distances are solved self-consistently as described in \textcolor{blue}{Section \ref{numerical}}. 

With this section we have completed the definition of all the physical quantities we use to calculate the dynamics of a subensemble of $n$ interacting polymers.

\section{Modeling the effective interchain potentials}
\label{potential}
The theory presented in this paper describes the temporal evolution of a subset of $n$ correlated macromolecules, where two distinct potentials operate between these chains: firstly, a many-body intermolecular potential of mean force acting between the center-of-mass of two polymers undergoing slow cooperative dynamics. This potential enters the zero mode of motion ($p=0$). Secondly, a potential that mimics the confinement due to entanglements, and impacts monomer dynamics through higher-order modes ($p=1, ..., N$) in the model.

\subsection{Many-body intermolecular potential of mean force}
In a liquid of neutral polymers, the effective intermolecular potential between chains reflects how local intermolecular monomer-monomer interactions propagate through the medium leading to the effective pair interactions between the center-of-mass of a pair of chains.\cite{Clark2012,Clark2013} The excluded volume intermolecular interaction between monomers generates an effective potential given by  the projection of these many-body interactions through the liquid onto a pair of effective sites, in this case the center-of-mass on each chain. The resulting potential has a dominant repulsive component at short distances and an attractive part at large distances. The attractive part is largely entropic in nature, due to the local degrees of freedoms that are averaged out during coarse-graining and the multiple liquid configurations.\cite{Clark2013,McCarty2014}

Given that the theory tracks the dynamics of a relatively small number of molecules, $n=\rho R_g^3/N$, undergoing slow cooperative motion, the effective potential is well-approximated by a potential of mean force, which is the potential between two molecules in the field of the others.\cite{Hansen2003,Guenza2018} The potential can be approximated by a Gaussian function, with a time-dependent spring constant for the center-of-mass intermolecular force given by\cite{Guenza1999,Guenza2002a,Guenza2002}
\begin{eqnarray}
\label{compot}
K_0[R_{cm}(t)]\approx - \frac{171}{32} \sqrt{\frac{3}{\pi}}\frac{\xi_{\rho}N}{R_g^3} k_BT
\Big(1 + \frac{\sqrt{2}\xi_\rho}{Rg} \Big) \ exp\Big[{-\frac{75 R_{cm}^2(t)}{76R_g^2}}\Big]   \ ,
\end{eqnarray}
with $R_{cm}^2(t) =\langle[\xi_0(t)-\xi_0(0)]^2\rangle/N$ the square intermolecular center-of-mass distance between a pair of molecules.\cite{Guenza1999} This distance evolves in time as polymers inter-diffuse, with  the  effective force being calculated at each time interval through a self-consistent procedure until the optimized intermolecular distance converges.

It's worth noting that the zero mode, which displays the center-of-mass subdiffusive motion, also influences monomer dynamics, as the monomer's coordinate is derived from the summation across all modes (Eq.\ref{monomerposition}). Consequently, cooperativity also impacts monomer mean-square displacement (see \textcolor{blue}{Figure \ref{monomercmmsd}}).

\subsection{Potential representing the constraint due to entanglement}

In the volume occupied by a chain, $R_g^3$, the chain is statistically in contact with $n-1$ other chains. Entanglements occur between the ``tagged'' chain and the $n-1$ other chains that are interpenetrating. The number of entanglements that the ``tagged'' chain experiences is given by the total number of monomers in the volume of the interpenetrating chains,  $(n-1)N \approx \rho R_g^3$, divided by the number of monomers in a chain segment between a pair of entanglements, $N_e$. The statistical number of entanglements per chain is then $(n-1)N/N_e\approx \rho R_g^3/N_e$.

\begin{figure}
\includegraphics[width=3.25in]{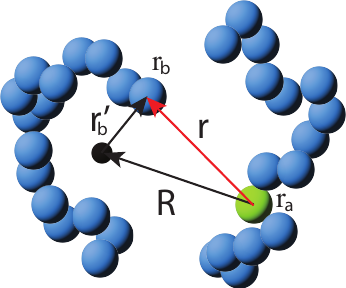} \\
\caption{Two monomers, $a$ and $b$, belonging to two different polymer chains are separated by a distance $r$. Given the distribution of a chain's monomers around the polymer's center-of-mass, the average distance between the monomer $a$ and a generic monomer $b$ is given by $R$. Here, $R$ is the average distance between the monomer $a$ and any monomer $b$ in the second polymer chain.}
\label{Model}
\end{figure}

Our model accounts for the effect of entanglements by applying a potential that is zero at any inter-monomer distance  smaller than a characteristic distance, $d$, which relates to the 'tube' diameter in the reptation picture. When two monomers, initially at contact, move a relative distance larger than the given average value, $d$, they experience an effective potential that tends to confine their relative motion to that distance. 
%The intermonomer potential due to entanglement is zero until the monomers reach a relative  distance comparable to $d$. 

%The confining potential is weighted by the probability of finding the two monomers inside two entangled polymers. 
To solve the potential, we start by recognizing that any monomer, $a$, within a polymer chain has a joint probability, $g(r)$, of encountering another monomer, $b$, belonging to a different chain at a distance $r=|\mathbf{r}_b - \mathbf{r}_a|$. 
If we define $\mathbf{R}$ as the distance between monomer $a$ in the first chain, and the average position of a generic monomer $b$ in the second chain (see \textcolor{blue}{Figure \ref{Model}}), then  $r=|\mathbf{r}|=|\mathbf{R} + \mathbf{r}'_b|$, with $|\mathbf{r}'_b|$ the average distance of a monomer $b$ from its polymer's center-of-mass. 
The potential of mean force between a pair of monomers, $a$ and $b$, belonging to two mutually entangled chains inside the correlation hole can be written as
$V[r(t)]=  - k_B T \rho R_g^3/N_e \ln [ g(r,t)]$, and the related intermolecular force constant is
\begin{eqnarray}
\label{forcec}
%\int d \Omega 
 K[r(t)] & \approx & - k_B T \rho R_g^3/N_e \langle \frac{1}{|r(t)|} \frac{\partial V[r(t)]}{\partial r(t)}  \rangle \nonumber \, \\ 
 & = &  k_BT \rho R_g^3/N_e \int d \mathbf{r}'_{b}(t) \frac{1}{|r(t)|}   \Psi(r'_b (t)) \frac{\partial g(r(t))}{\partial r}  \  ,
\end{eqnarray}
%where $ H[r-d] $ is the Heaviside step function.
where the potential is calculated as an average over the position distribution of the monomer $b$ in the second polymer,
\begin{eqnarray}
\Psi(r'_b (t))=\Big( \frac{3}{2 \pi Rg^2} \Big)^{3/2} e^{-\frac{3}{2}\frac{\mathbf{r}'_b(t)^2}{Rg^2}} \ .
\end{eqnarray}

An analytical expression for the effective force acting between  the two monomers can be derived using the thread model representation of the monomer pair distribution function for a liquid of polymer chains. In the thread model a polymer is described as an infinitely thin and infinitely long chain, while the density of the liquid is kept constant.  In this model, the PRISM theory\cite{schweizer_prism_1994} gives for the pair distribution function of the monomers inside the volume defined by an entangled segment the simple analytical expression
\begin{eqnarray}
g(r)=1+\frac{3}{\pi \rho \sigma^2} \Big[  \frac{e^{-|\mathbf{r}|/ \xi_{\rho}}}{|\mathbf{r}|}  -  \frac{e^{-|\mathbf{r}|/ \xi_{d}}}{|\mathbf{r}|}   \Big] \ ,
\end{eqnarray}
which describes the joint probability of finding another monomer belonging to another polymer at some distance $|\mathbf{r}|$.
Here, $\xi_{\rho} \approx 3/\pi \rho \sigma^2$ is the local density fluctuation screening length, which is related to the liquid packing fraction and the bulk properties of the systems, such as the liquid compressibility. The second characteristic lengthscale is the entanglement length $\xi_d \approx d/\sqrt{2}$. For distances larger than the 'tube' diameter, each monomer has a high probability of encountering another polymer, as each chain is confined within the 'tube' and does not move outside of it. 

Under these approximations, the equation of the force constant can be solved analytically, as a function of the average distance between the monomer $a$ and a generic monomer $b$ in the second chain, $R(t)$, as
\begin{eqnarray}
\label{forceconstant}
K[R(t)] = k_B T/N_e \frac{3 \sqrt{3}}{\sqrt{2 \pi}} \frac{3}{\pi \sigma^2} \ln{\frac{\xi_d^2}{\xi_{\rho}^2}} e^{-\frac{3 R(t)^2}{2 Rg^2} } \ ,
\end{eqnarray}
with the derivation of Eq.\ref{forceconstant} presented in the \textcolor{blue}{Section \ref{AppendixA}}. The confining potential is then given by $V[R(t)]=0$ if $R(t) \le d $ and
\begin{equation}
\label{entpot}
V[R(t)] \approx 
k_B T/N_e \frac{3}{\pi \sigma^2}  e^{-3 R(t)^2/2R_g^2} \  [R(t)-d]^2\ ,
\end{equation}
for $R(t) > d$.

\begin{figure}
\includegraphics[width=3.25in]{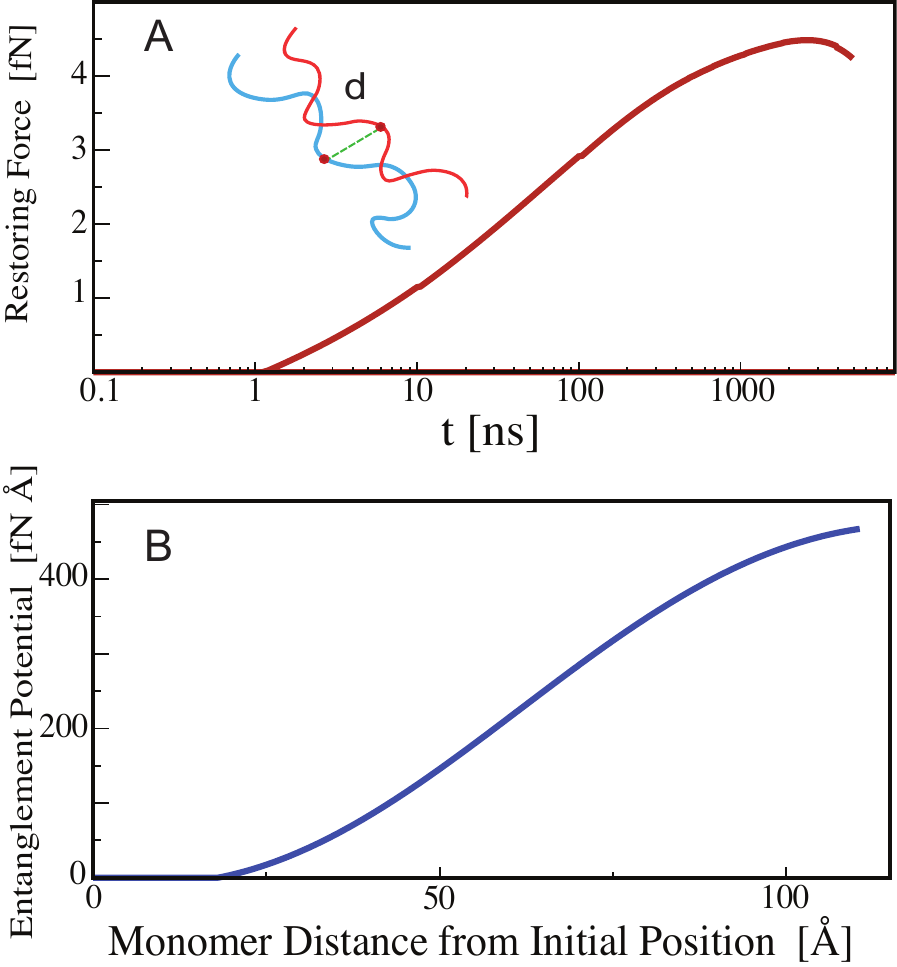} \\
\caption{\textbf{(\textit{A})}: Effective force that a monomer experiences as a function of time when it moves a distance comparable to $R(t)\approx d$ with respect to another monomer belonging to another chain entangled with the first. The two monomers, which are initially in contact, inter diffuse as a function of time and as they move apart they experience an increasing confining force. This force could be represented by an effective "tube" in a single-chain formalism.  \\
\textbf{(\textit{B})}: Effective entanglement potential between two monomers belonging to two entangled chains. The shape of this curve depends on the thermodynamic conditions and on the chemical structure of the polymer.}
\label{Figure1}
\end{figure}

At any give time, $t$, the potential is harmonic. However, the time dependence of the spring constant leads to an effective time-dependent potential that is anharmonic, as illustrated in \textcolor{blue}{Figure \ref{Figure1}}.

Harmonic potentials have been used in the literature to describe the confinement of entanglements. Slip-link models are numerical methods used to simulate the dynamics of a single entangled polymer chain, with the entanglements represented as harmonic constraints.\cite{Schieber1998,Likhtman2014} Slip-link models are single-chain, largely phenomenological, and anisotropic. In contrast, our potential is isotropic and anharmonic, which limits the application of the CDGLE in its present form to unentangled and weakly entangled systems.  More detailed discussions of the strengths and limitations of slip-link models are present in the literature.\cite{SchC2SM26674A}

The force constant enters Eq.(\ref{lambdadn}) and applies at the monomer level to both unentangled and weakly entangled polymers. While it has a relevant impact on the dynamics of long chains, it has no effect on the dynamics of short chains because their motion becomes uncorrelated before they move an intermolecular monomer distance comparable to $d$. The numerical value of the parameter $d$ is obtained from fitting the NSE experimental data. Once the optimal value of $d$ is defined, the equation of motion is solved self-consistently as described in \textcolor{blue}{Section \ref{numerical}}.

\begin{figure}
\includegraphics[width=3.25in]{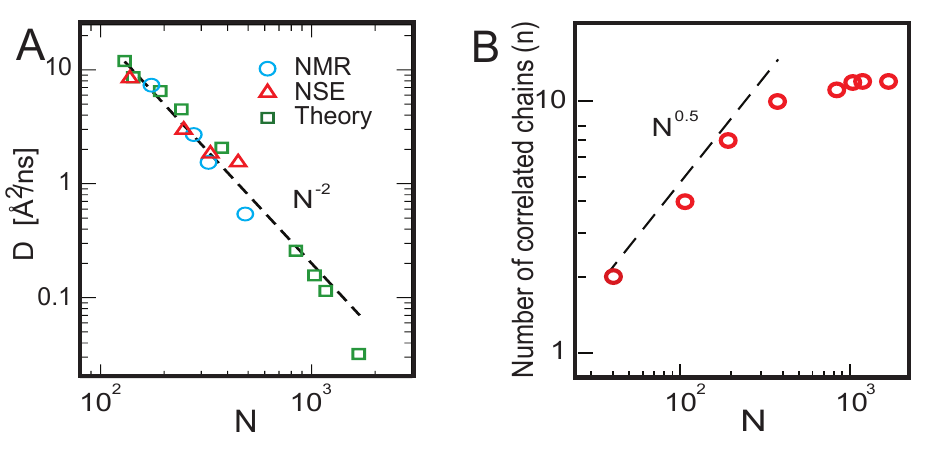} \\
\caption{\textbf{(\textit{A})}: 
 Diffusion coefficient of entangled samples, calculated using the numerically optimized friction coefficient (green squares) as a function of the degree of chain polymerization, and compared with NMR data from Ref. \cite{Pearson2002a,Pearson2002} (light blue circles), and NSE data from Ref. \cite{Richter1993} (red triangle). The dashed line depicts the scaling of $N^{-2}$, typical of entangled polymers.
\textbf{(\textit{B})}: Number of chains undergoing correlated dynamics ($n$). This parameter is optimized by fitting the CDGLE theory to the NSE experiments. The figure shows that while the number of correlated chains increases as the square root of the degree of polymerization in unentangled polymer samples ($N \langle 130$), it remains constant for samples that are entangled ($N \rangle 130$). See text for more details.}
\label{Figure2}
\end{figure}

\textcolor{blue}{Panel \textit{A}} of \textcolor{blue}{Figure \ref{Figure1}} displays the time-dependent effective force that confines a monomer due to the presence of entanglements, obtained from the self-consistent solution of the CDGLE equation of motion, while \textcolor{blue}{Panel \textit{B}} of \textcolor{blue}{Figure \ref{Figure1}} shows the related potential.
The force grows smoothly
as the distance between the two monomers increases. It shows confinement that opposes the free diffusion of the monomer and increases as the monomer approaches the point of entanglement.
The range and intensity of the confining force and potentials are in agreement with simulations\cite{Zhou2006,Ramanathan} and experiments.\cite{Robertson2007,Wang2010,Glaser2010} Notably, unlike slip-link models that use phenomenological harmonic potentials to constrain the dynamics, the effective potential in this approach is time-dependent and anharmonic. The confining force and potential depend through Eq.\ref{forceconstant} on the sample's density and temperature, on the polymer radius of gyration and effective segment length, as well as on the entanglement distance $d$. The distance $d$ is the only free parameter optimized to reproduce the NSE data, while the other parameters are determined by the type of polymers and the thermodynamic conditions of the sample.

It is noteworthy that the different polyethylene samples in this study, which have increasing degree of polymerization and whose parameter $d$ is optimized \textit{independently}, show identical values of the optimized $d$ parameter and identical confinement force. The onset of the confining force when two close monomers separate by the distance $d$ occurs at about $100$ ns (\textcolor{blue}{Panel \textit{A}} in \textcolor{blue}{Figure \ref{Figure1}}): this time window agrees with the crossover of the center-of-mass subdiffusive to diffusive dynamics for entangled chains in the chain mean-square-displacement (see \textcolor{blue}{Panel \textit{B}} in \textcolor{blue}{Figure \ref{monomercmmsd}}). Within the subdiffusive region ($t < 100 ns $), the center-of-mass motion is dominated by the "tube confinement", and different polymer chains are equivalently slowed down. However, the crossover to diffusive motion increases with chain length because diffusion decreases with $N$. The crossover time is called, in reptation theory, the disengagement time or $\tau_d$. 

\section{Methods: numerical solution of the CDGLE}
\label{numerical}
\subsection{Self-consistent solution of the CDGLE with entanglements}
\label{selfconsistent}
In this section we explain the numerical self-consistent procedure that is used to solve the extended CDGLE equation, Eq.\ref{eq:langevin}. The zero mode potential depends on the relative distance of the center-of-mass of two chains, which changes with time as the chains interdiffuse. Thus, Eq.\ref{drcm} is solved self-consistently at a fixed time interval until we obtain agreement between the assumed distance  between the centers-of-mass of a pair of chains and the distance predicted by solving the equation of motion. Initially, the procedure adopts a small interpolymer distance, selecting an ensemble of chains close in space and interacting. As the calcolation proceeds the polymers interdiffuse away from each other and they dynamics ultimately become uncorrelated.

Once the inter-polymer distance is optimized for a fixed time interval, we optimize the average interchain monomer-monomer distance following a similar procedure. The potential that constraints the relative motion of entangled chains (Eq.\ref{entpot}) acts between pairs of monomers belonging to two different polymers in the sub-ensemble of chains that are interpenetrating at initial time. Being local, this potential enters the monomer equations of motion, Eq.\ref{eq:coordinates0} with Eq.\ref{tdsc}, which are solved self-consistently for the intermonomer distance, Eq.\ref{mondist}.

After both distances have converged, the system moves an infinitesimal time-step forward and the whole procedure is repeated having as initial guess for the distances the values calculated in the preceding time step. This convergence procedure is quite robust and is not sensitive to small differences in the length of the time interval selected, or to the chosen initial values adopted for the distances, so long as the time step is short enough at initial times. 

At each given time, after optimizing the intermolecular center-of-mass and monomer distances, we calculate the effective forces that contribute to the CDGLE time correlation functions, as well as the structural and dynamical quantities of interest. These include the new intermolecular distances, which are then optimized self-consistently until convergence is achieved.

\subsection{Effective parameters entering the theory}
\label{numericalparameters}
The  Langevin Equation for cooperative dynamics requires a number of molecular and physical parameters as an input. Most parameters are defined either from the analysis of computer simulations or from the experiments. 

The experiments define the molecular parameters (polymer flexibility as polyethylene's persistence length or the $g$ parameter for the freely-rotating-chain model, and degree of polymerization, $N$) and the thermodynamic parameters (temperature, $T$, and monomer density, $\rho$). The non-Gaussian parameter, $\alpha$  defined below is calculated from simulations performed for polyethylene at the given thermodynamic conditions. From the same simulations we calculated the semiflexibility parameter, $g= - \langle \cos \theta \rangle$ of the freely-rotating-chain model, which agrees with the one reported in the literature for polyethylene, $g=0.785$. These simulations were previously documented in our papers and will not be extensively discussed here.\cite{Dinpajooh2018a,Guenza2018} Briefly, we completed a set of  LAMMPS\cite{Plimpton1995} MD simulations of polyethylene represented by United Atoms (UA-MD) at the same thermodynamic conditions of the experimental data, with the temperature $T=509 \ K$, and the density $\rho=0.733$ $g/cm^3$,\cite{Schleger1998} and increasing degree of polymerization from $N=16$ to $N=300$ monomers.  All the simulations were performed in the canonical ensemble using the Nose-Hoover thermostat, following the procedure described in our previous papers.\cite{Dinpajooh2018a,Guenza2018} 

Other parameters entering the theory are parameters that we obtain from fitting the data from the NSE experiments as briefly described below. Those parameters are the entanglement length, $d$, the number of correlated chains, $n$, and the monomer friction coefficient, $\zeta_{eff}$.

The distance $d$ entering the confining potential,(Eq.\ref{entpot} and \textcolor{blue}{Figure\ref{Figure1}}), is equivalent to the statistical length of the chain segment between two entanglements, $d=\sqrt{N_e} |\mathbf{l}|$. Experimentally, the value of $N_e$ has been traditionally measured using a variety of  methods such as melt rheology, neutron spin echo, and NMR relaxation.  Different methods may lead to different values of $N_e$.\cite{Lodge1999} 
This discrepancy  may result from the fact that there are differences in the theoretical models used to interpret the data.\cite{Likhtman2005} Note that $N_e$ can be also obtained from an analysis of computer simulations using the primitive path method\cite{Everaers2012}, the Contour REduction Topological Analysis algorithm or CRETA algorithm,\cite{Tzoumanekas2009}, the Z-code and others.\cite{Hoy2009} These methods give slightly different values of $N_e$ for polyethylene and  the entanglement distance $d$ is estimated to be somewhat smaller than the experimental one. 

In our study the calculation of $d$ by direct comparison of the self-consistent calculation with the data of neutron spin echo gives an identical outcome for all the entangled samples even if the parameter is evaluated for each sample independently. The values that optimizes the agreement with NSE experiments ($d=17 \ \AA$) is consistent with $d=\sqrt{N_e} l$ for $l=1.53 \ \AA$ and $N_e\approx 130$ for polyethylene.

Another input parameter is the monomer effective friction coefficient which is obtained in this study using two different procedures depending on the degree of polymerization of the samples. \textit{Short} chains, which follow unentangled dynamics, have fast relaxation and reach the region of Fickian dynamics during the timescale of the NSE experiments (see \textcolor{blue}{Figure \ref{monomercmmsd}, Panel \textit{B}}). For these NSE samples the monomer friction coefficient is calculated from the diffusion coefficient in the region for $t > \tau_{decorr}$ for unentangled chains 
where the chain mean-square-displacement scales linearly in time, as $\zeta_{eff}=k_BT/(N D)$, with $k_B$ the Boltzmann constant. 
%The regime of Brownian motion is also sampled well in the UA-MD simulations which provide complementary data to the experiments.

The friction coefficient of \textit{long} entangled chains is more difficult to derive from NSE experiments because those samples do not reach the diffusive regime in the timescale of the experiments. For these samples, we calculate the friction by optimizing the agreement between the theory and  the dynamic structure factor of the experiments. The resulting diffusion coefficients, which are reported in \textcolor{blue}{Figure \ref{Figure2}, Panel \textit{A}}, for the entangled polymers are consistent with the values obtained from independent experiments of NMR \cite{Pearson2002a,Pearson2002} and NSE.\cite{Richter1993} They also follow the scaling with $N$ expected for long chains with this number of entanglements. 
The values of friction coefficients are also consistent with the diffusion coefficients we calculated using a reconstruction procedure from the mesoscale simulations of coarse-grained polyethylene melts.\cite{Lyubimov2011,Lyubimov2010} 

The last non-trivial parameter in our calculation is the number of chains undergoing correlated dynamics, $n$. Approximatively, one can estimate the value of $n$ using the number of chains that statistically occupy the volume defined by the range of the intermolecular potential, i.e., the polymer correlation hole. The range of the potential, for samples at the density and temperature of the NSE data, is of the order of the polymer radius of gyration, $R_g=\sqrt{N}l_{eff}/6$. Given that for melts, the monomer density $\rho=nN/R_g^3\approx 1$, the number of correlated chains is $n\approx \rho \sqrt{N} l_{eff}^3$. For chains of fixed length, $N$, the number of correlated chains is expected to increase with increasing chain stiffness as the overall volume spanned by one chain increases. It also increases with increasing chain density. 
Thus, for chains of melts of homogeneous composition and constant density, the number of chains undergoing cooperative dynamics should grow as $ n \propto N^{0.5}$.

In this study, as previously mentioned, the number of chains undergoing cooperative dynamics is treated as an adjustable parameter, the value of which is determined through optimization to directly align the theoretical predictions with experimental observations. We find that $n$ increases as $N^{0.5}$ for unentangled and slightly entangled chains in agreement with the theoretical predictions, while it  becomes fairly constant in the entangled regime (see \textcolor{blue}{Figure \ref{Figure2}, Panel \textit{B}}). Thus, our study suggests that the number of chains undergoing cooperative motion grows until the crossover to entangled dynamics, where the cooperative dynamics becomes confined to the region between entanglements, and $n\approx \rho \sqrt{N_e} l_{eff}^3$, or, to use a picture close to the `reptation' model, in entangled polymers the cooperative motion, and the related sub-diffusive behavior, is dominated by the presence of entanglements, which confine the correlated motion of the chain to the region inside the 'tube'.\cite{Guenza2014} Thus, the dynamics in segments shorter than the entanglement length can still show subdiffusive behaviour due to the cooperative motion of the interprenetrating chains inside the correlation hole, while larger segments of the chain are confined by the entanglements and display subdiffusive motion only on the local scale.

%~~~~~~

\section{Results: Comparison of the CDGLE with Neutron Spin Echo experiments for weakly entangled and unentangled polymers}
\label{NSE}
In this section we present the CDGLE predictions for mean-square displacements and dynamic structure factor. We compare the theoretical predictions with Neutron Spin Echo data of polyethylene samples at the temperature 
$T=509$ $K$, monomer density $\rho=0.733$ $g/cm^3$, and increasing degree of polymerization, $N=192$, $N=377$, $849$, $1041$, $1178$, and $1692$.\cite{Schleger1998} The input parameters to the CDGLE approach are presented in \textcolor{blue}{Section \ref{numerical}}. 

\subsection{Mean-square displacements}
\label{msds}
The incoherent part of the intermediate scattering function measured in NSE experiments is defined as 
\begin{eqnarray}
\label{Skt}
S(\mathbf{k},t)= N^{-1} \sum_{i,j=1}^N \langle e^{i \mathbf{k} \cdot [\mathbf{r}_i(t) - \mathbf{r}_j(0)]}\rangle \ ,
\end{eqnarray}
where we drop the chain index $a$ because the monomers $i$ and $j$ belong to the same polymer. Here, $S(\mathbf{k},t)$ can be expressed as the exponential function of an infinite series in the momentum transfer $k^2$.\cite{Kubo1966,Rahman1962}  In the condition under which the distribution of displacement is Gaussian, only even contributions are relevant and the $k^2$ term is the leading one,\cite{Zorn1997} 
\begin{eqnarray}
\label{Skt1}
S(\mathbf{k},t)= N^{-1} \sum_{i,j=1}^N e^{- \frac{k^2}{6}\langle[\mathbf{r}_i(t) - \mathbf{r}_j(0)]^2\rangle} \ ,
\end{eqnarray}
and is calculated in the CDGLE approach from the intramolecular time-dependent monomer-monomer distance of Eq.\ref{perskt}, which includes contributions due to cooperativity, entanglements, and chain semiflexibility.

\begin{figure}
\includegraphics[width=3.25in]{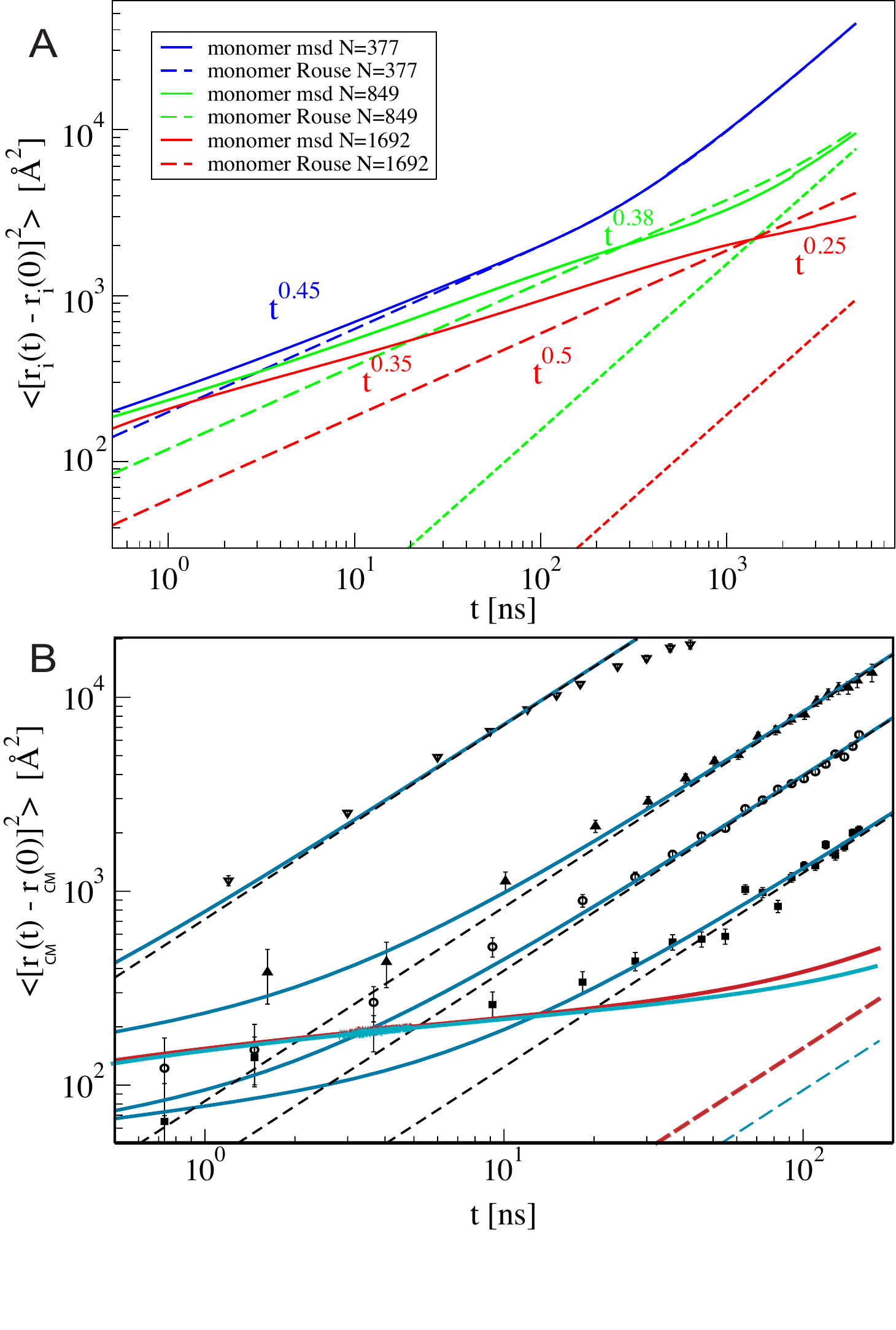} \\
\caption{\textbf{(\textit{A})} Monomer mean square displacement for polyethylene with $N=377$, $N=849$, and $N=1962$ (from top to bottom). The CDGLE formalism (full line) is compared with the Rouse predictions for a fully flexible chain (dashed lines). The Rouse equation scales in time following $t^\nu$ with $\nu=0.5$ exponent, while the CDGLE shows the crossover from cooperative motion of a semiflexible chain ($\nu \langle0.5$) to the reptation characteristic exponent of $\nu=0.25$ in the long time region. \\
 \textbf{(\textit{B})}: Center-of-mass mean square displacement as a function of time for polyethylene melts of increasing chain length. Symbols are data from Neutron Spin Echo, superimposing lines are the theory presented here. From top to bottom, $N=36$ (triangle down), $106$ (triangle up), $192$ (circle), $377$ (squares), $N=1041$ (red line) and $N=1178$ (light blue line). For samples $N=1041$ and $N=1178$, experimental points are not reported (see text). Long-time Fickian dynamics is represented by the dashed lines. }
\label{monomercmmsd}
\end{figure}

The self contribution in the exponent of the dynamic structure factor, Eq.\ref{Skt1}, is the mean-square monomer displacement calculated using Eq.\ref{eqmonomermsd}.
\textcolor{blue}{Figure \ref{monomercmmsd}} illustrates the monomer mean square displacement predicted by the CDGLE for some of the NSE samples studied here. Specifically, it shows the mean square displacement, calculated using Eq.\ref{eqmonomermsd}, for polyethylene with $N=377$, $849$, and $1962$. These samples represent the crossover from unentangled to entangled dynamics, covering a range of entanglements from two entanglements for $N=377$ to twelve entanglements for $N=1692$. The CDGLE formalism for a semflexible chain undergoing cooperative dynamics is compared with the Rouse predictions for a fully flexible chain. The Rouse model scales with time following $t^\nu$, with an exponent $\nu=0.5$ as expected. In contrast, the CDGLE shows that, a short times, the cooperative motion of a semiflexible chain is characterized by an exponent smaller than that of the Rouse motion ($\nu < 0.5$). In the short-time regime, the monomer dynamics become slower with increasing chain length ($\nu \approx 0.45$, $0.38$, and $0.35$ for $N=377$, $849$, and $1692$, respectively). At long times, the monomer displacement transitions to center-of-mass diffusion for the shorter chain. For the entangled samples, it crosses to the reptation motion, with the characteristic exponent $\nu=0.25$, before transitioning to the expected diffusive dynamics.

In the low wavevector regime, where $k R_g << 1$,
\begin{eqnarray}
\lim_{k \rightarrow 0} \ln \frac{S(\mathbf{k},t)}{N} \approx \frac{k^2}{6} \langle[\mathbf{r}_{cm}(t) - \mathbf{r}_{cm}(0)]^2\rangle  \ .
\end{eqnarray}
Consequently, in the case of short chains where $R_g$ is small, one can compute the mean-square displacement from the low $k$ limit of $S(\mathbf{k},t)$.  \textcolor{blue}{Panel \textit{B}} of \textcolor{blue}{Figure \ref{monomercmmsd} } illustrates the center-of-mass mean-square displacement versus time, including both the data from Neutron Spin Echo (NSE) and the theoretical predictions of the extended CDGLE: the two are in excellent agreement. As we previously mentioned, in the case of entangled chains where the radius of gyration is large, the values of the momentum transfer, $k$, reported in the experimental NSE  data are not low enough to fulfill the required limit. Therefore, it is impossible to directly obtain the center of mass diffusion from the NSE data in such scenarios. For those entangled chains we report in our figure only the theoretical predictions. 

\textcolor{blue}{Panel \textit{B}} of \textcolor{blue}{Figure \ref{Figure1} } shows that, consistently with the known dynamical behavior of polymer melts, the MSD of the center of mass follows diffusive dynamics at long time, $\Delta R(t)^2 \propto t$. The most interesting behavior emerges at shorter time scales, where the dynamics is sub-diffusive, with $\Delta R(t)^2 \propto t^{\nu}$ and $\nu < 1$, for all the samples, including short, unentangled chains. The subdiffusive behavior becomes increasingly more pronunced with increasing degree of polymerization. Note that, akin to undercooled systems, the subdiffusive behavior observed in the molecular center-of-mass serves as an indication of heterogeneous dynamics and correlated cooperative motion in this context.\cite{Guenza2022} 

In the CDGLE  theory, the correlation of the dynamics due to the intermolecular intermonomer potential-of-mean-force leads to a slowing down of the center-of-mass motion as long as the chains are within the range of the effective mean-field potential. 
The subdiffusive dynamics persists for longer time for longer unentangled chains, given that the range of the potential of mean force is consistent with the polymer correlation hole and hence with $R_g$. Because the polymer must diffuse beyond the potential's range for its dynamics to lose correlation, the transition to Brownian motion occurs at a distance comparable to the unentangled polymer's size.

The picture is slightly different for entangled chains where we find that entanglements dominate the subdiffusive motion of the chains. For very long chains the subdiffusive dynamics is terminated when entanglements' dynamics set in and the center-of-mass mean-square displacement displays subdiffusive motion only inside the volume defined by the entanglement length scale, $d$. Furthermore, inside that volume the subdiffusive dynamics is identical, and involves the same number of chains in the cooperative motion, for all the entangled samples (see \textcolor{blue}{Panel \textit{B}} of \textcolor{blue}{Figure \ref{monomercmmsd}}).

\subsection{Dynamic structure factor}
\label{SKTt}
The incoherent scattering function, Eq.\ref{Skt1}  depends on the distance between two monomers belonging to the same polymer  
$\langle[\mathbf{r}_i(t) - \mathbf{r}_j(0)]^2\rangle$, which is calculated following Eq.\ref{perskt}.
For dynamically heterogenous systems, like ours, 
the incoherent scattering function includes a $q^4$ correction term to the Gaussian approximation,
$ln\langle e^{i \mathbf{k} \cdot \Delta \mathbf{r}(t)}\rangle \approx - \frac{k^2}{6}\langle[\Delta \mathbf{r}(t)]^2\rangle + \frac{k^4}{72}\langle[\Delta \mathbf{r}(t)]^2\rangle^2  \alpha_2(t) - O(k^6)$. 
This term is a small contribution to the overall dynamic structure factor.\cite{Guenza2014} For all the NSE samples available, the CDGLE predictions are in very good agreement with NSE experiments. Moreover, to analyze further the components of the observed structure factor's decay, we take advantage of our  analytical formalism to evaluate the shape and weight of each contribution of semiflexibility, cooperativity, and entanglements to the overall polymer dynamics for chains of increasing length in the melt. 

\begin{figure}
\includegraphics[width=3.25in]{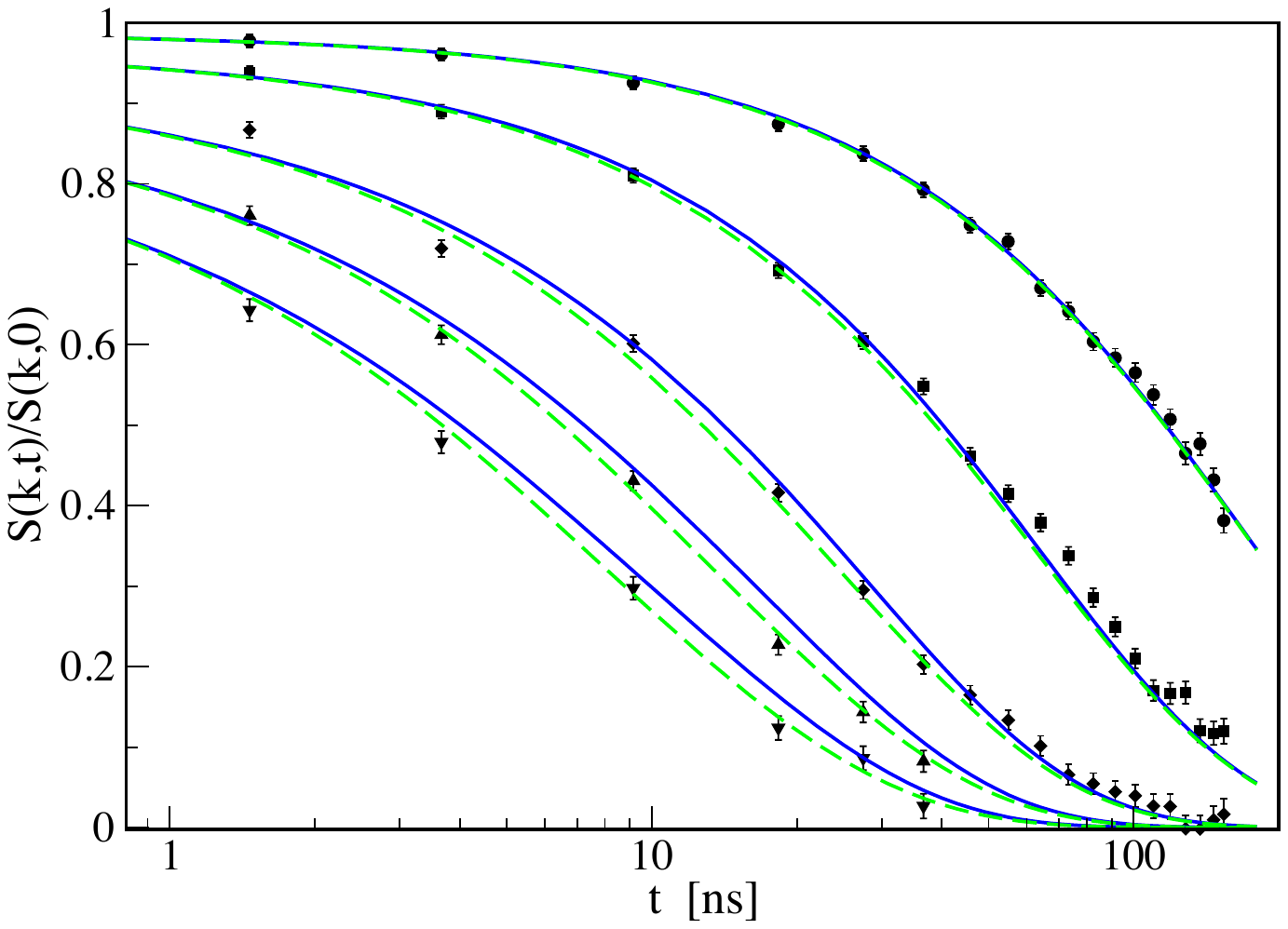} \\
\caption{Comparison between  theoretical (lines) and experimental (symbols) values of the normalized incoherent intermediate scattering function for polyethylene with $N=192$ with (full lines) and without (dashed lines) the contribution due to entanglements. For this short chain the number of entanglements is equal to one, and entanglement effects on the dynamics are small. Data are 
at increasing wave vector $q=0.3$ (circle), $0.5$ (square), $0.77$ (diamond), $0.96$ (triangle up), $1.15$ (triangle down). }
\label{figure3}
\end{figure}

\begin{figure}
\includegraphics[width=7.0in]{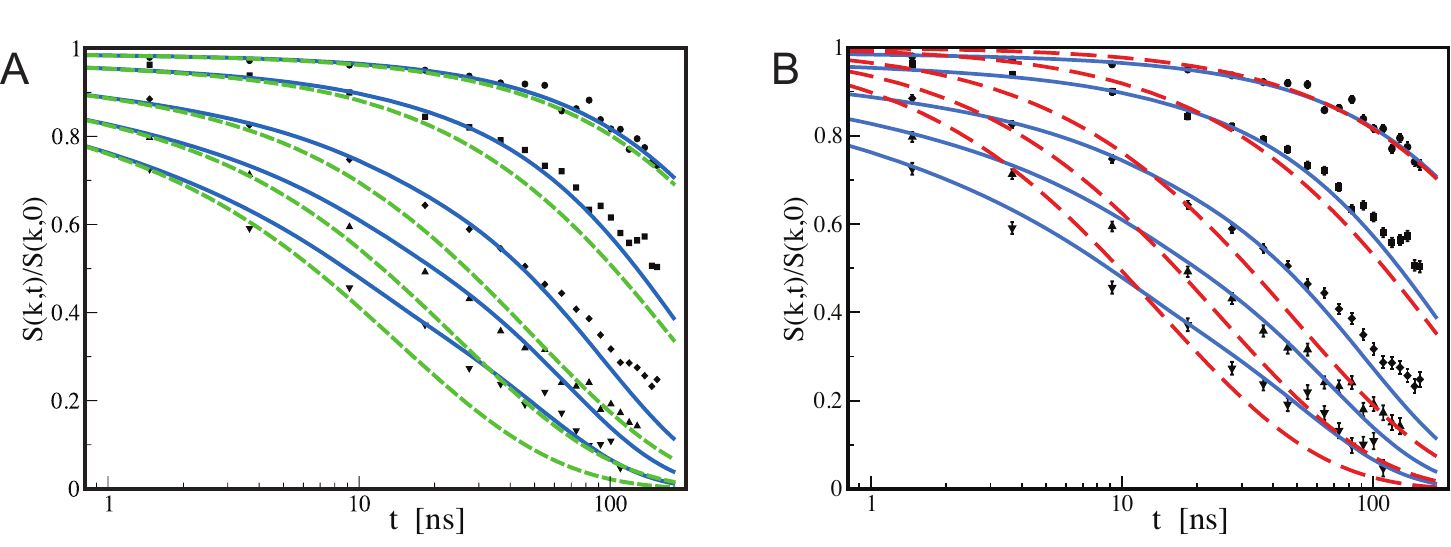} \\
\caption{\textbf{(\textit{A})}:Comparison between  theoretical (lines) and experimental (symbols) values of the normalized incoherent intermediate scattering function for polyethylene with $N=377$ with entanglement (full lines) and without entanglements (dashed lines). Data are 
at increasing wave vector $q=0.3$ (circle), $0.5$ (square), $0.77$ (diamond), $0.96$ (triangle up), $1.15$ (triangle down).\\
\textbf{(\textit{B})}:
Comparison between  theoretical (lines) and experimental (symbols) values of the normalized incoherent intermediate scattering function for polyethylene $N=377$ (full lines) and the Rouse dynamics for a flexible chain of equivalent mean-square end-to-end distance (dashed lines). Data are 
at increasing wave vector $q=0.3$ (circle), $0.5$ (square), $0.77$ (diamond), $0.96$ (triangle up), $1.15$ (triangle down). }
\label{Figure4}
\end{figure}

\textit{Contributions due to entanglements.} In \textcolor{blue}{Figure \ref{figure3}} and in \textcolor{blue}{Panel \textit{A}} of \textcolor{blue}{Figure \ref{Figure4}},  experimental data are compared with two different versions of the theory: the CDGLE where the contribution from entanglements is accounted for, and the CDGLE theory where entanglements are neglected, meaning that the intermonomer potential is zero. In both figures the inclusion of entanglements improve the agreement with experiments. 
The CDGLE formalism in both cases describes semiflexibile chains modeled as FRCs, using a Bixon-Zwanzig type of approach as presented in \textcolor{blue}{Section \ref{backgrounddynamics}}.\cite{Bixon1973,Bixon1977}

Note however, that the sample with $N=192$ in \textcolor{blue}{Figure \ref{figure3}} has statistically only one entanglement, and its dynamics is well described also by the Langevin Equation for cooperative dynamics with no entanglements.\cite{Zamponi2008} However, while the difference between the two curves is small, the correction due to the presence of the entanglements still improves the agreement of the theory with the data.

The entanglement effect is more pronounced in the $N=377$ sample in 
\textcolor{blue}{Figure \ref{Figure4} Panel \textit{A}}, where including the entanglement potential in the CDGLE improves the long-time decay of the structure factor and its agreement with the data in the long-time scale. 
In this sample, each chain in the melt has statistically two entanglements. Thus polyethylene with $N=377$ is at the crossover region between unentangled and entangled dynamics. While similar effects are observed in all the samples, the effect of entanglements becomes more pronounced as the number of entanglements per chain increases.

\textit{Cooperativity and semiflexibility.} To quantify the combined effects of cooperativity, semiflexibility, and entanglements we map the freely-rotating-chain chain in the CDGLE onto a fully flexible chain represented by a freely-jointed chain model, where the "beads" are connected by effective bonds of length $l_{eff}=\sqrt{17} \ \AA$. This mapping ensures that the freely jointed chain of $N$ segments of length $l_{eff}$ has the same mean-square end-to-end distance of the freely-rotating-chain of $N$ segments of length $l$, correlated by an effective semiflexibility parameter of $g=0.785$. 
By solving our CDGLE formalism while enforcing $g=0$, $l_{eff}$, and $n=0$, we calculate the dynamics of a chain completely flexible, unentangled, and uncorrelated with the surrounding chains, i.e. the dynamics of a Rouse chain, where the chain has the same number of "beads" and equivalent mean-square end-to-end distance than the sample.

In the reptation model of entangled dynamics, the motion inside the "tube" is represented by the Rouse model. The most relevant region of S(k,t) where the Rouse dynamics should apply is the short time region of the plot and the large wave-vector, $k$, region where the effects of the entanglements are not yet felt by the polymer. In \textcolor{blue}{Figure \ref{Figure4} Panel \textit{B}}, we compare the dynamics of the CDGLE chain with experimental data of the incoherent scattering function and with the Rouse formalism for a polyethylene sample with $N=377$ beads. While the CDGLE agreement with the experiments is excellent in the whole range of time scales, the Rouse model is unable to reproduce, not even qualitatively, the observed experimental decay. This is because the Rouse model is missing both cooperativity and local semiflexibility. The dynamical correlation observed in the mean-square center-of-mass displacement of \textcolor{blue}{Figure \ref{Figure1}, Panel \textit{B}} as subdiffusive motion is also visible in the short-time decay of the dynamic structure factors for all the NSE samples.

\begin{figure}
\includegraphics[width=7.0in]{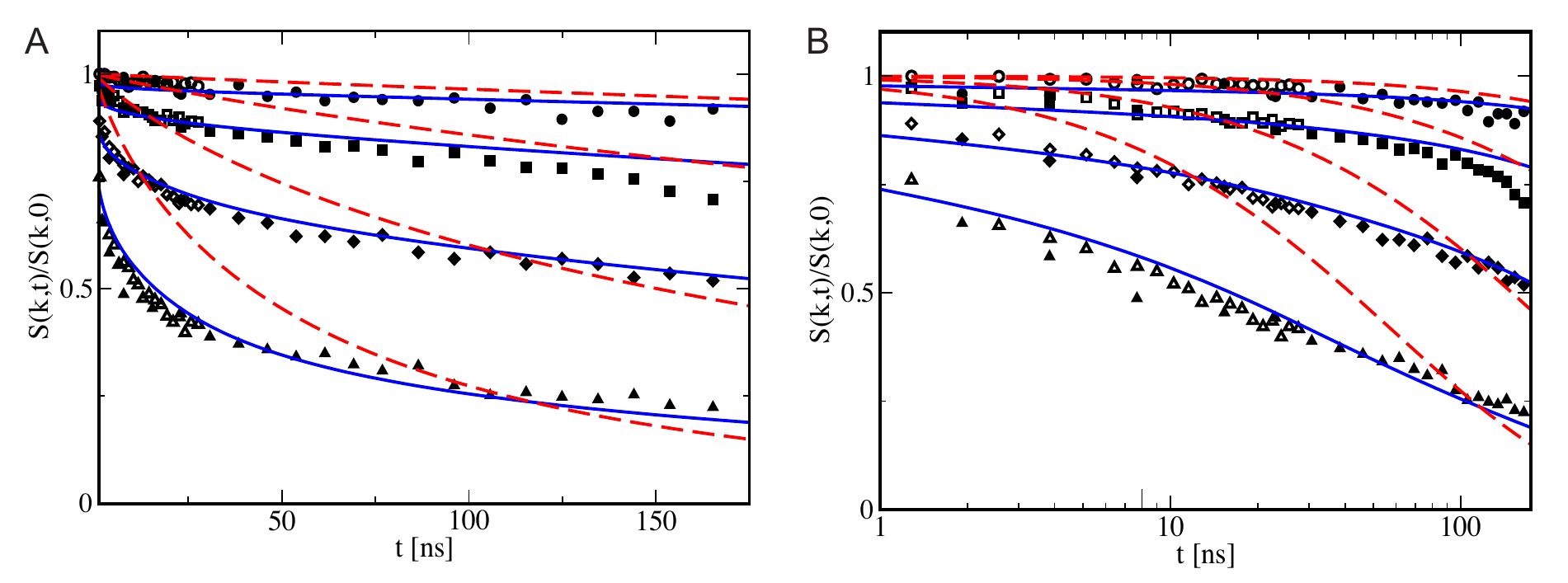} \\
\caption{\textbf{(\textit{A})}:Comparison between  theoretical (lines) and experimental (symbols) values of the normalized incoherent intermediate scattering function for polyethylene with $N=849$ including entanglement (full lines) and the Rouse dynamics of a flexible chain without entanglements and cooperative dynamics (dashed lines) Data are 
at increasing wave vector $q=0.3$ (circle), $0.5$ (square), $0.77$ (diamond), $1.15$ (triangle up).\\
\textbf{(\textit{B})}:
Same as in Panel \textit{A}, but reported on a logarithmic time scale to emphasize the agreement between CDGLE theory and experiments in the short-time regime.}
\label{figure4a}
\end{figure}

\textcolor{blue}{Figure \ref{figure4a}} presents a comparison between the CDGLE approach and the NSE data for $N=849$, alongside theoretical predictions for a semiflexible chain without entanglements and with cooperative motion. The figure illustrates excellent agreement both in the short-time regime, emphasized by the logarithmic scale, and in the long-time regime, evident on the linear scale.

\begin{figure}
\includegraphics[width=3.25in]{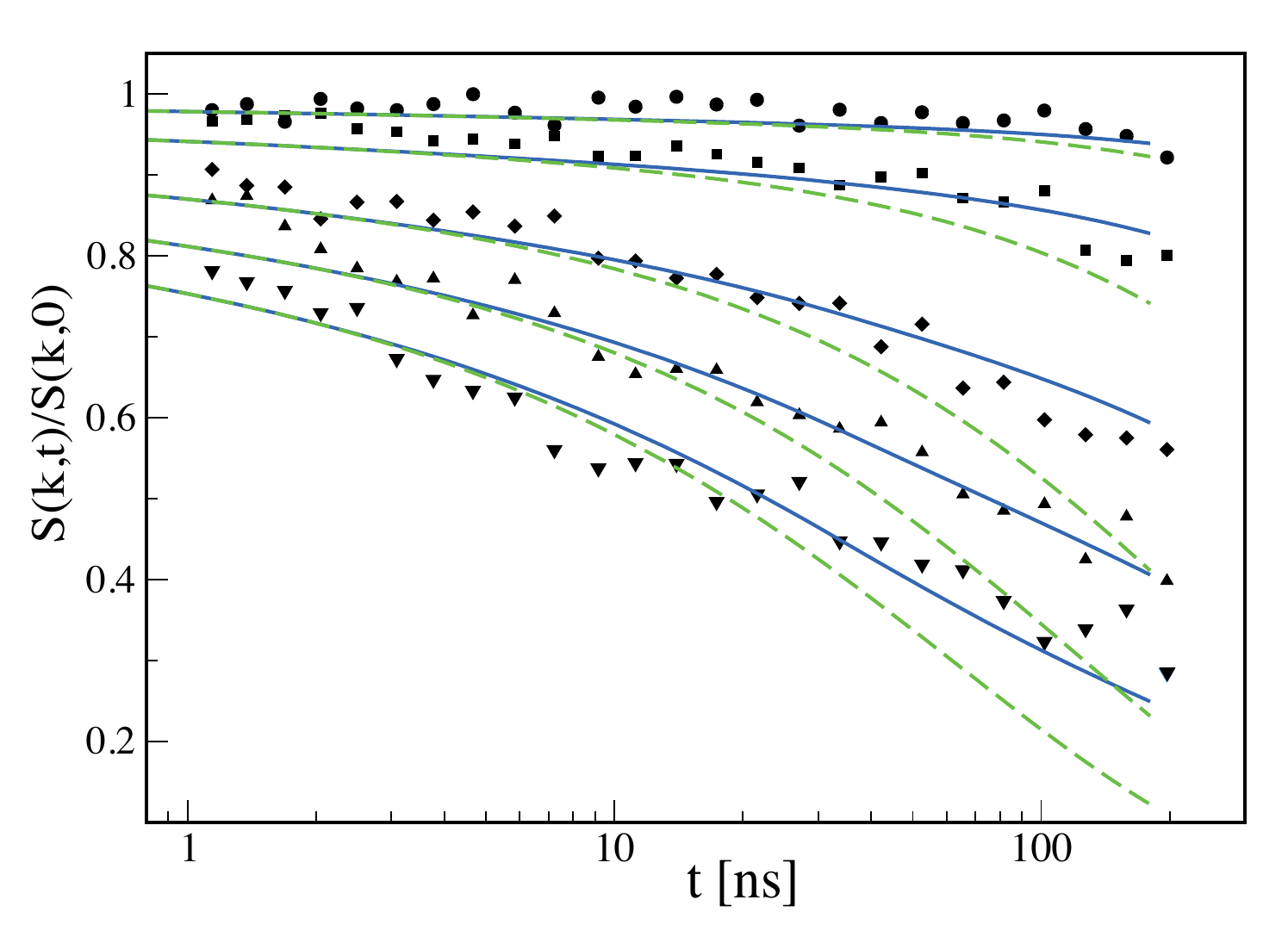} \\
\caption{Comparison between  theoretical (lines) and experimental (symbols) values of the normalized incoherent intermediate scattering function for polyethylene $N=1041$ with (full lines) and without (dashed lines) the contribution due to entanglements. Data are 
at increasing wave vector $q=0.3$ (circle), $0.5$ (square), $0.77$ (diamond), $0.96$ (triangle up), $1.15$ (triangle down). }
\label{figure5}
\end{figure}

\begin{figure}
\includegraphics[width=3.25in]{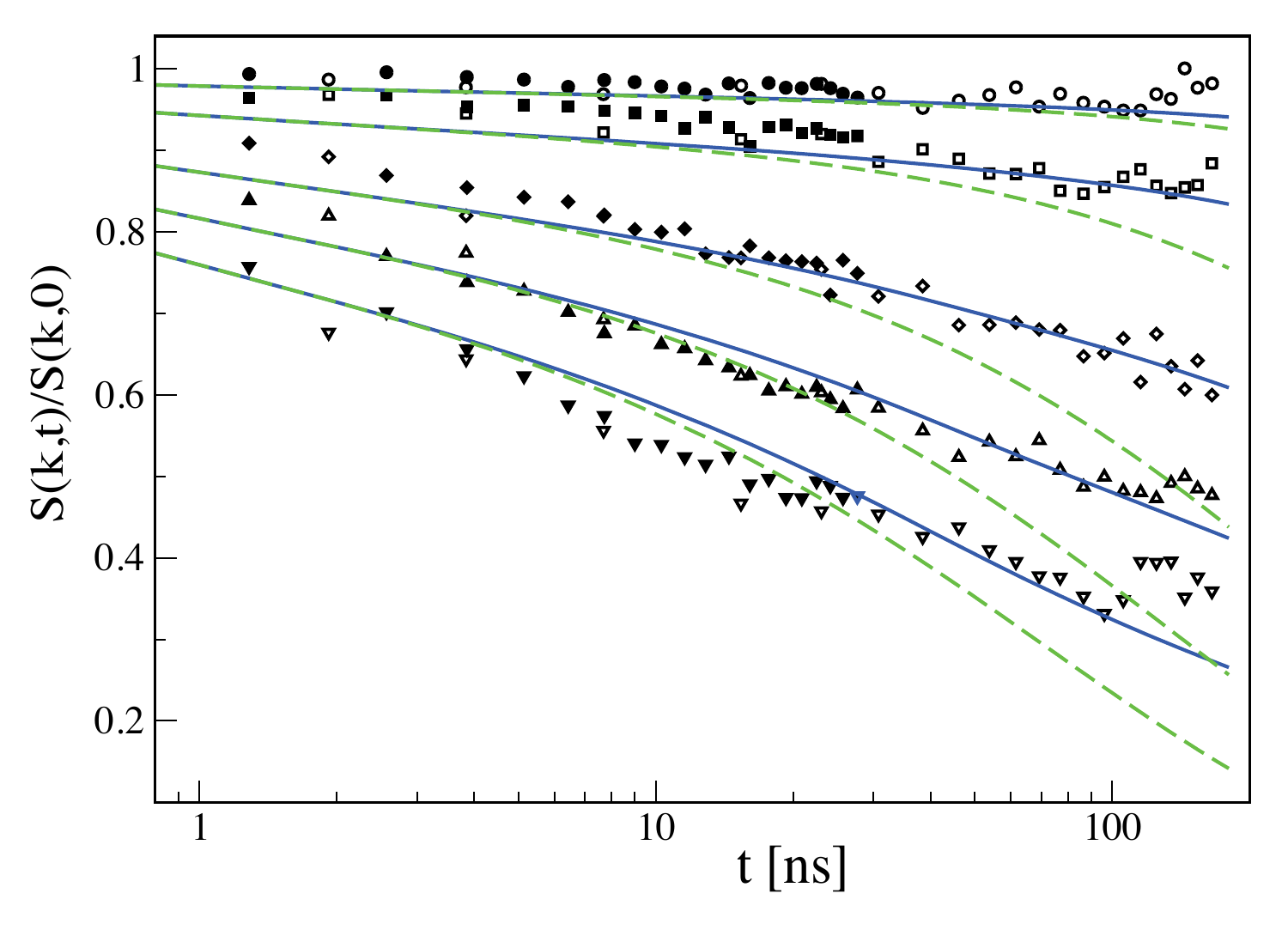} \\
\caption{ Comparison between  theoretical (lines) and experimental (symbols) values of the normalized incoherent intermediate scattering function for polyethylene $N=1178$ with (full lines) and without (dashed lines) the contribution due to entanglements. Data are 
at increasing wave vector $q=0.3$ (circle), $0.5$ (square), $0.77$ (diamond), $0.96$ (triangle up), $1.15$ (triangle down).}
\label{figure7}
\end{figure}

\begin{figure}
\includegraphics[width=3.25in]{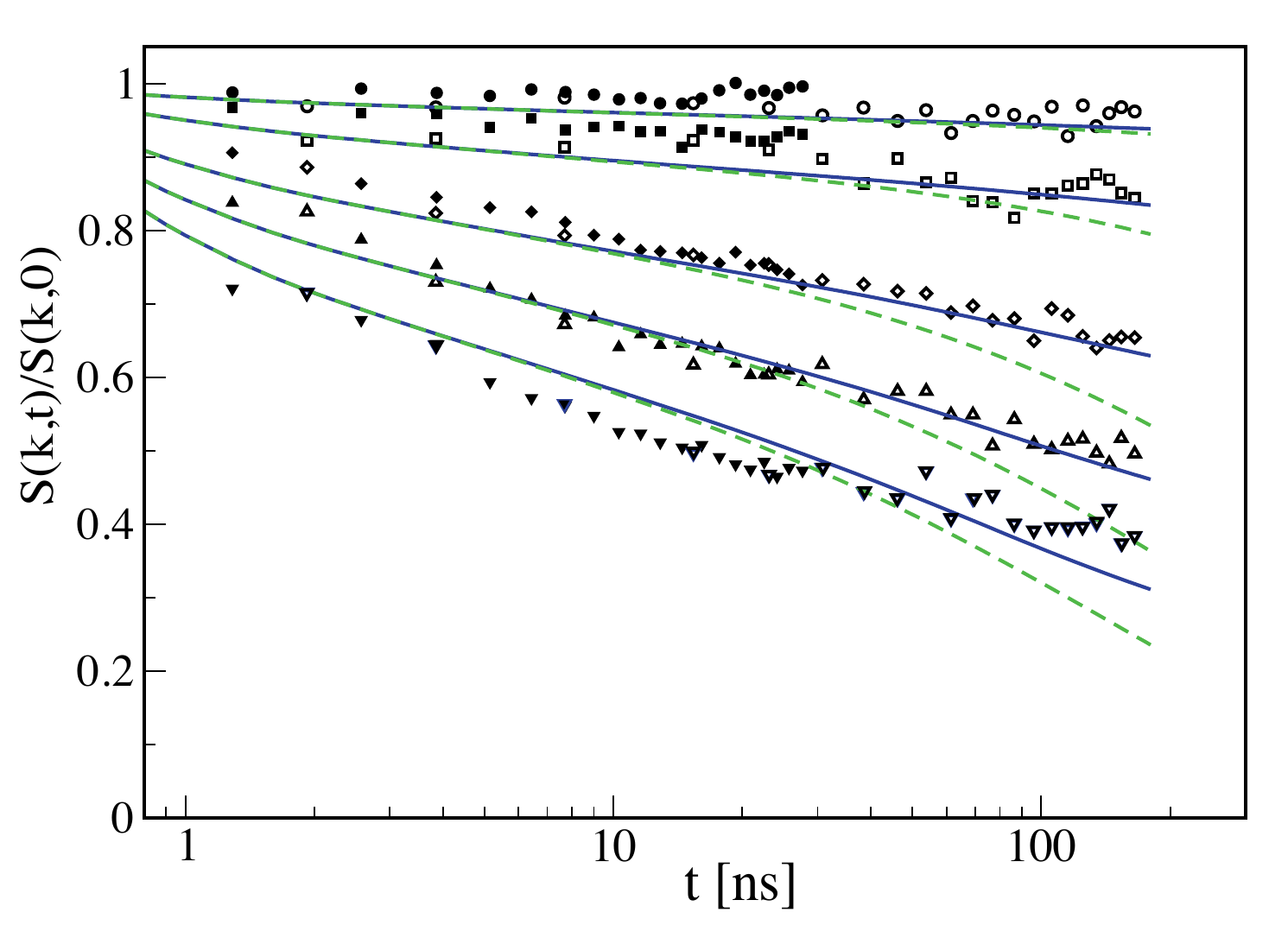} \\
\caption{ Comparison between  theoretical (lines) and experimental (symbols) values of the normalized incoherent intermediate scattering function for polyethylene $N=1692$ with (full lines) and without (dashed lines) the contribution due to entanglements. Data are 
at increasing wave vector $q=0.3$ (circle), $0.5$ (square), $0.77$ (diamond), $0.96$ (triangle up), $1.15$ (triangle down).}
\label{figure8}
\end{figure}

The following three figures offer a comparison between CDGLE theory and NSE experiments for weakly entangled polyethylene chains, together with CDGLE calculations excluding entanglements. This comparison elucidate the impact of entanglements on chain dynamics. In \textcolor{blue}{Figure \ref{figure5}}, experimental data for $N=1041$ are compared with the theory where semiflexibility and cooperativity of the dynamics are accounted for but entanglements are not included. The theory for semiflexible unentangled chains agrees well with the experiments in the short time region. Similar plots are presented for chains with a higher number of entanglements, namely $N=1178$ corresponding to $8$ entanglements and $N=1692$ corresponding to $12$ entanglements. Those plots are shown in \textcolor{blue}{Figure \ref{figure7}} and in \textcolor{blue}{Figure \ref{figure8}}, respectively.

In all the plots the theory without the presence of entanglements agrees well with the experiments in the short time region, for $t \le 4 \ ns$. This general behavior is expected considering the fact that at short time the monomer dynamics is not affected by the length of the polymer chain the monomer belong to, and only samples its local environment. For the center-of-mass motion, the dynamics inside the entangled region is dominated by cooperativity, which is similar for all entangled chains, given that the motion is very local. The concept that the dynamics on a local scale is common to all the polymers that have different chain lengths but identical chemical structure agrees with the  physical hypothesis that motivates Ngai's Coupling Model approach, where  local and global dynamics connect at an intermediate time scale which is similar for all degrees of polymerization of the polymer samples.\cite{Ngai2011}

\section{Cooperative dynamics in the shear relaxation}
\label{shear}
\subsection{Theory: CDGLE stress relaxation under linear shear perturbation}
The effect of cooperative dynamics on dynamical mechanical measurements of a polymer liquid under linear shear flow is described by the CDGLE approach. The complete derivation is presented in \text color{blue}{section \ref{AppendixB}}.\cite{Bird1987,Doi1988} Briefly, the stress tensor for a sub-ensemble of chains undergoing cooperative dynamics under a linear shear flow defined as $v_x=\dot{\gamma}(t) \ y$, with $\dot{\gamma}(t)$ the shear rate matrix, is given by  
\begin{eqnarray}
\label{sigma}
\sigma_{\alpha,\beta}(t)= \frac{\rho}{nN} k_s \sum_{i,j=1}^N \Big[(n-1) \langle r^D_{i,\alpha}(t) A^D_{i,j} r^D_{j,\beta}(t)\rangle + \langle r^N_{i,\alpha}(t) A^N_{i,j} r^N_{j,\beta}(t)\rangle\Big]   \ , 
\end{eqnarray}
with $\alpha=x$ and $\beta=y$, where 
the tensor is partitioned into a relative and a collective contribution.

After applying the transformation into normal modes, the general expression for the stress tensor is 
\begin{eqnarray}
\label{stresstensora}
\sigma_{\alpha,\beta}(t)= \frac{\rho}{N} k_B T  \Bigg[4 (n-1)^2  \int_0^t d\tau \frac{K_0[r(t)]}{K_0[r(\tau)]} e^{-\frac{2n}{\zeta_{eff}}\big[K_0[r(t)] - K_0[r(\tau)]\big]} \ \dot{\gamma}(\tau) \  + \sum_{p=1}^{N-1} \int_0^t d\tau e^{-\frac{2 k_s \Lambda^D_p (t-\tau)}{\zeta_{eff}}} \ \dot{\gamma}(\tau)  \Bigg] \ , 
\end{eqnarray}
which gives for the CDGLE shear relaxation modulus of a group of interacting polymer chains undergoing cooperative dynamics  
\begin{eqnarray}
\label{gta}
  G(t)= \frac{\rho}{N} k_B T  \Bigg[4 (n-1)^2 e^{-\frac{R^2(t)}{R_g^2}} e^{-\frac{2n}{\zeta_{eff}}\int_0^t K_0[r(t')]dt'} + \sum_{p=1}^{N-1} e^{-\frac{2 k_s \Lambda^D_p (t)}{\zeta_{eff}}}\Bigg] \ .
\end{eqnarray}
Note that the CDGLE predicts a more complex decay of the shear relaxation modulus than the straightforward multi-exponential decay described by Rouse. Here intermolecular correlations in the dynamics of interacting chains  significantly influence the modulus's relaxation through the contribution of the zero mode. 

This is the final CDGLE equation for which we report some model calculations in the following section.

\subsection{Results: CDGLE predictions of shear relaxation in polyethylene melts with increasing degrees of polymerization}
Polyethylene is a polymer known for its propensity to  crystallize, which poses challenges for measuring shear relaxation under the thermodynamic conditions and time window of the NSE experiments studied in this paper.\cite{Ferry1980} Consequently, we present the CDGLE calculations of shear relaxation to study the effect of cooperativity and semiflexibility on the dynamical mechanical properties of a melt comprising semiflexible polymers of increasing degree of polymerization. We adopt the same parameters of the samples previously investigated by NSE and reported in this paper.\cite{Schleger1998,Zamponi2008}

\begin{figure}
\includegraphics[width=7.0in]{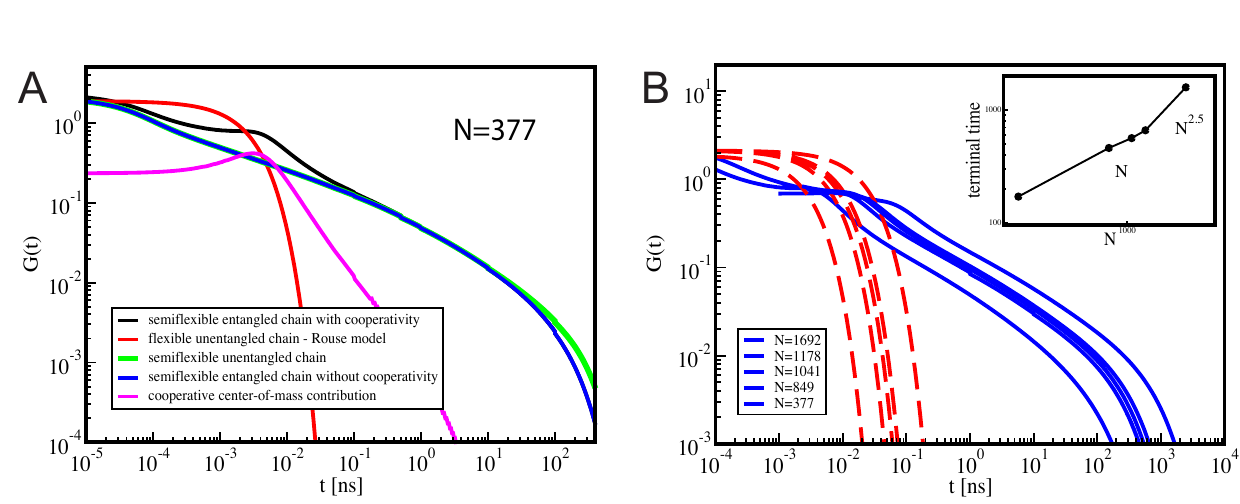} \\
\caption{\textbf{(\textit{A})}:CDGLE predictions of the shear relaxation for polyethylene $N=377$. The figure shows the contributions to the total relaxation. The presence of semiflexibility slows down the relaxation, while the effect of cooperativity leads to a shoulder visible in the short-time regime.  
\textbf{(\textit{B})}: 
comparison of the shear modulus for chains of increasing length, $N=377$, $N=849$, $N=1041$, $N=1178$, and $N=1692$. The decay of $G(t)$ occurs on timescales that increase with the degree of polymerization, as expected. The dashed lines are the shear modulus predicted by the Rouse model for flexible unentangled chains with those degrees of polymerization. The insert shows the scaling with $N$ of the terminal time.}
\label{figureG(t)}
\end{figure}

In \textcolor{blue}{Figure \ref{figureG(t)}}, we illustrate the predictions of the CDGLE for shear relaxation of a polyethylene chain within a subset of interacting polymers undergoing cooperative dynamics. The samples have varying degrees of polymerization.\textcolor{blue}{Panel \textit{A}} shows model calculations of $G(t)$ for a sample with $N=377$, where the polymers are modeled as semiflexible FRC chains, both with and without entanglements. The cooperative dynamics contribution to $G(t)$ appears as a small plateau in the short-time region of the plot, characterized by a rapid decay of the function. The short-time contribution to shear relaxation, arising from cooperativity, is intermolecular and decays over the time scale at which subdiffusive dynamics transition to diffusive behavior in the center-of-mass mean-square displacement. \textcolor{blue}{Panel \textit{B}} highlights that shear modulus relaxation requires increasingly longer times as the degree of polymerization of the weakly entangled chains increases. The inset illustrates the terminal time determined at a fixed $G(t)$ value for samples of increasing $N$. The terminal time is calculated for samples of increasing chain length by identifying the time at which each curve crosses the $G(t)=10^{-3}$ line. This is an approximate estimate of $\tau=\int_0^\infty G(t)$, based on the assumption that $G(t)\approx e^{-t/\tau}$. By imposing that $G(t)_{N=1692}=G(t)_{N=1178}$ and so on for all the samples, it defines the time for which $t_{N=1692}=\tau_{N=1692}$, $t_{N=1178}=\tau_{N=1178}$, and so on for all the samples. The integral of $G(t)$ is the shear viscosity. While for shorter chains, this terminal time adheres to a scaling akin to Rouse scaling, the longest chain exhibits a transition to much slower times, seemly scaling with $N$ with an approximate exponent of $2.5$. Because this sample is only weakly entangled, its dynamics does not conform yet to fully entangled scaling, nor does it show the expected plateau in $G(t)$.\cite{Doi1988} Testing the entanglement scaling behavior of viscosity within this polymerization range is not feasible.

\section{Solution of the entanglement force}
\label{AppendixA}
Because the intermolecular force experienced by the molecules depends on their reciprocal  distance, the entanglement force evolves with time.\cite{Guenza1999} Eq.\ref{eq:langevin} is solved through a self-consistent procedure at a fixed time interval; fluctuations of the intermolecular distance are allowed, but the average intermonomer distance is at each time step the one that is obtained from the optimization. Because the equation is solved at a fixed time interval, we drop in our notation the time dependence  (see Eq.\ref{forcec}) with the understanding that the intermonomer distance is in itself time dependent. 

The equation of the force has two type of terms that are related by a simple transformation
\begin{eqnarray}
\label{secondterm}
\frac{1}{r^2}e^{-r/\xi}= -\frac{\partial }{\partial \xi^{-1}}\frac{e^{-r/\xi}}{r^3} \ ,
\end{eqnarray}
 so that the only integral to solve is of the type
 \begin{eqnarray}
 I=- \beta^{-1} \int d \mathbf{r}_{b} \Psi(r_b) \frac{e^{-r/\xi}}{r^3} \ .
\end{eqnarray} 
    
By introducing the Fourier transforms
\begin{eqnarray}
\frac{1}{|\mathbf{r}|}= -\frac{4 \pi}{(2 \pi)^3}\int d\mathbf{k} \frac{e^{i \mathbf{k} \cdot \mathbf{r}}}{k^2} \ ,
\end{eqnarray}
and 
\begin{eqnarray}
\frac{e^{-|\mathbf{r}|/\xi}}{|\mathbf{r}|}= -\frac{4 \pi}{(2 \pi)^3}\int d\mathbf{k} \frac{ \xi^2e^{i \mathbf{k} \cdot \mathbf{r}}}{1+k^2 \xi^2} \ ,
\end{eqnarray}
the integral is rewritten as
 \begin{eqnarray}
 \label{integrale}
I  = - \beta^{-1}  \ \frac{1}{8 \pi^6}  \int \frac{d \mathbf{k}_1}{k_1^2}  \int \frac{d \mathbf{k}_2}{k_2^2}  
\int \frac{d \mathbf{k}_3}{\xi^{-2}+ k_3^2} 
 \int d \mathbf{r}_b  e^{i \mathbf{k} \cdot \mathbf{r}} \ e^{ - \mu  {r'_{b}}^2} \ ,
\end{eqnarray}
where $\mathbf{k}=\mathbf{k}_1 + \mathbf{k}_2 + \mathbf{k}_3$, $\mu=3/(2 R_g^2)$, and $\mathbf{r}=\mathbf{R}+\mathbf{r}'_b$. 

Next, we introduce the change of variables 
\begin{eqnarray}
\mathbf{k'} & = &\mathbf{k}_1+\mathbf{k}_2 \nonumber \\
\mathbf{k}_3 & = &\mathbf{k}-\mathbf{k'}\\
\mathbf{k}_2 & = &\mathbf{k'}-\mathbf{k}_1 \nonumber \ ,
\end{eqnarray}
which leads to 
\begin{eqnarray}
I  = - \beta^{-1}  \ \frac{1}{8 \pi^3} \int \frac{d \mathbf{k'}}{\mathbf{k'}} \int \frac{d \mathbf{k}}{\xi^{-2}+|\mathbf{k}-\mathbf{k'}|^2} e^{i \mathbf{k} \cdot \mathbf{R}} \int d \mathbf{r}'_b \ e^{i \mathbf{k} \cdot \mathbf{r}'_b} \ e^{ - \mu {r'_{b}}^2}
\end{eqnarray}
given the convolution integral 
\begin{equation}
\int d \mathbf{k}_1\frac{1}{\mathbf{k}_1^2 |\mathbf{k'}-\mathbf{k}_1|^2}= \frac{(2 \pi)^3}{8 \mathbf{k'}} \ .
\end{equation}
Using the solution
\begin{equation}
\int d \mathbf{k'}\frac{1}{\mathbf{k'} (\xi^{-2}+ |\mathbf{k}-\mathbf{k'}|^2)}= - \frac{(4 \pi)}{k}\Bigg[\frac{arccot(k\xi)}{\xi} - \frac{k}{2} \ln{(k^2+\xi^{-2})}  \Bigg] \ ,
\end{equation}
and solving the second integral with Eq.\ref{secondterm}, the $arccot$ term cancels. Because the  exponential term is dominant at $k\approx N^{-1}$, where $k^2\xi_d^2\approx d/N$ and $k^2\xi_{\rho}^2\approx N^{-1}$, in that range $\ln{(k^2+\xi^{-2})}\approx -\ln{\xi^2}$.
Then, the entanglement force constant is well approximated as 
\begin{eqnarray}
K[R(t)] = k_B T/N_e \frac{3 \sqrt{3}}{\sqrt{2 \pi}} \frac{3}{\pi \sigma^2} \ln{\frac{\xi_d^2}{\xi_{\rho}^2}} e^{-\frac{3 R(t)^2}{2 Rg^2} } \  ,
\end{eqnarray}
which is Eq.\ref{forceconstant}.

\section{Solution of the CDGLE stress relaxation under linear shear perturbation}
\label{AppendixB}
We now turn to the calculations of the effect of cooperative dynamics on dynamical mechanical measurements, specifically for a polymer liquid under linear shear flow.\cite{Bird1987,Doi1988}

The stress tensor for a sub-ensemble of chains undergoing cooperative dynamics under a linear shear flow defined as $v_x=\dot{\gamma}(t) \ y$, with $\dot{\gamma}(t)$ the shear rate matrix, is given by  
\begin{equation}
\sigma_{\alpha,\beta}(t)= - \frac{\rho}{nN}\sum_{i,j=1}^{nN} \langle F_{i,\alpha}(t) r_{j,\beta}(t)\rangle \ ,
\end{equation}
with $\alpha=x$ and $\beta=y$. 
By inserting the similarity transformation matrix,
the tensor is partitioned into a relative and a collective contribution,
\begin{eqnarray}
\label{sigma}
\sigma_{\alpha,\beta}(t)= \frac{\rho}{nN} k_s \sum_{i,j=1}^N \Big[(n-1) \langle r^D_{i,\alpha}(t) A^D_{i,j} r^D_{j,\beta}(t)\rangle + \langle r^N_{i,\alpha}(t) A^N_{i,j} r^N_{j,\beta}(t)\rangle\Big]   \ . 
\end{eqnarray}
After transforming the stress tensor into relative and collective diffusive modes, we disregard the first collective mode. This mode represents the diffusion of a cluster of interacting chains and does not contribute to the shear stress, thereby simplifying Eq.\ref{sigma} to:
\begin{eqnarray}
\label{sigmamode}
\sigma_{\alpha,\beta}(t)= \frac{\rho}{nN} k_s \Big[(n-1)\sum_{p=0}^{N-1} \Lambda^D_p \langle \xi_{p,\alpha}(t) \xi_{p,\beta}(t)\rangle + \sum_{p=1}^{N-1} \Lambda^N_p \langle\chi_{p,\alpha}(t) \chi_{p,\beta}(t)\rangle\Big]   \ ,
\end{eqnarray}
where the center-of-mass relative diffusion of a group of interacting chains still enters the tensor. Here, the eigenvalues $\Lambda^D_p$ and $\Lambda^N_p$ for mode numbers $p=1, ..., N-1$ are defined in Eq.\ref{lambdadn}, while the eigenvalue $\Lambda^D_0$, corresponding to mode $p=0$, is defined in Eq.\ref{lambda0}.

For the \textit{internal modes}, for which $p=1, 2, ..., N-1$, the cross correlation functions in the relative coordinates are 
\begin{eqnarray}
 \langle\xi_{p,\alpha}(t)\xi_{p,\beta}(t)\rangle = \frac{k_BT}{k_s \Lambda^D_p} \sum_{p=1}^{N-1} \int_0^t d\tau e^{-\frac{2 k_s \Lambda^D_p (t-\tau)}{\zeta_{eff}}} \dot{\gamma}(t)   \ , 
\end{eqnarray}
and similarly for the collective coordinates. Thus, the total contribution to the shear from the internal coordinates is
\begin{eqnarray}
\label{sigmamode1}
\sigma_{\alpha,\beta}(t)_{p=1,2, ...,N-1}=  \frac{k_BT \rho}{N} \sum_{p=1}^{N-1} \int_0^t d\tau e^{-\frac{2 k_s \Lambda^D_p (t-\tau)}{\zeta_{eff}}} \dot{\gamma}(t) \ ,
\end{eqnarray}
which gives for the shear relaxation modulus  
\begin{eqnarray}
\label{modulus}
    G(t)= \frac{k_BT \rho}{N} \sum_{p=1}^{N-1} e^{-2k_s \Lambda^D_p t} \ .
\end{eqnarray}
By applying the conditions of no cooperative dynamics ($n=1$) and no entanglements, along with complete polymer flexibility ($g=0$), it is straightforward to derive from Eq. \ref{modulus} the well-known Rouse expression for the modulus. In fact, under these conditions, $\Lambda^D_p$ equals $\lambda_{p}^{Rouse}$.

However, the CDGLE predicts a more complex decay of the shear relaxation modulus than the straightforward multi-exponential decay described by Rouse. This is due to the presence of cooperative chain motion, where intermolecular correlations in the dynamics of interacting chains  significantly influence the modulus's relaxation through the contribution of the zero mode. For the \textit{center of mass relative} mode, $\Lambda^D_0$ is given by $n K_0[r(t)]/k_s$, whereas for the collective mode, $\Lambda^N_0$ equals zero. The stress tensor for the zero mode is as follows:
\begin{eqnarray}
\label{sigmamodea}
\sigma_{\alpha,\beta}(t)_{p=0}= \frac{\rho}{nN} (n-1) n K_0[r(t)] \langle \xi_{p,\alpha}(t) \xi_{p,\beta}(t)\rangle   \ ,
\end{eqnarray}
with time evolution under shear of the $\alpha$ and $\beta$ components of the zero mode 
\begin{eqnarray}
\label{crosszeroa}
%\langle\xi_{0,\alpha}(t)\xi_{0,\beta}(t)\rangle & = & \langle\xi_{0,\alpha}(0)\xi_{0,\beta}(0)\rangle e^{-\alpha t} +  e^{-\alpha t} \int_0^t d\tau \frac{k_B T}{ n K_0[r(t)]}  e^{\alpha \tau} \ , \\
\langle\xi_{0,\alpha}(t)\xi_{0,\beta}(t)\rangle & = & 4 k_B T (n-1) \int_0^t d\tau \frac{e^{-\frac{2n}{\zeta_{eff}}\big[K_0[r(t)] - K_0[r(\tau)]\big]}}{K_0[r(\tau)]} \dot{\gamma}(\tau) \ .
\end{eqnarray}

The general expression for the stress tensor is:
\begin{eqnarray}
\label{stresstensora}
\sigma_{\alpha,\beta}(t)= \frac{\rho}{N} k_B T  \Bigg[4 (n-1)^2  \int_0^t d\tau \frac{K_0[r(t)]}{K_0[r(\tau)]} e^{-\frac{2n}{\zeta_{eff}}\big[K_0[r(t)] - K_0[r(\tau)]\big]} \ \dot{\gamma}(\tau) \  + \sum_{p=1}^{N-1} \int_0^t d\tau e^{-\frac{2 k_s \Lambda^D_p (t-\tau)}{\zeta_{eff}}} \ \dot{\gamma}(\tau)  \Bigg] \ ,
\end{eqnarray}
which gives for the shear relaxation modulus of a group of interacting polymer chains undergoing cooperative dynamics  
\begin{eqnarray}
\label{gta}
  G(t)= \frac{\rho}{N} k_B T  \Bigg[4 (n-1)^2 e^{-\frac{R^2(t)}{R_g^2}} e^{-\frac{2n}{\zeta_{eff}}\int_0^t K_0[r(t')]dt'} + \sum_{p=1}^{N-1} e^{-\frac{2 k_s \Lambda^D_p (t)}{\zeta_{eff}}}\Bigg] \ . 
\end{eqnarray}
This is the final CDGLE equation for which we report some model calculations in the following section.

\section{Conclusions}
This paper presents the theoretical formalism that extends the CDGLE approach to the dynamics of polymer melts covering systems with different degrees of polymerization, spanning the dynamics from the unentangled to the weakly entangled regime. The formalism doesn't need to be modified to describe the two different dynamical regimes observed for short and for long chains. Signatures of the crossover to entangled dynamics are observed in the monomer mean-square displacement, diffusion coefficient, and shear relaxation modulus for the sample with the highest degree of polymerization studied here. The proposed CDGLE approach is found to be in excellent agreement with NSE data, while predicting reasonable values of the fitting parameters, that are the friction coefficient coefficient and the number of chains undergoing cooperative dynamics. 

The flexibility of the formalism in addressing different systems is due to the many-chain nature of the approach. 
The theory selects at initial time a group of interpenetrating chains interacting through an effective center-of-mass pair potential that represents the projection of the many-body monomer-monomer interactions onto the polymer center-of-mass of a pair of chains. The monomer-monomer interactions propagate through the liquid of macromolecules surrounding the slow-moving chains undergoing cooperative motion. 
%The mean field contribution due to the units that are projected out during the coarse-graining of the dynamics, which leads to the Langevin equation, result in an effective potential which is long-ranged and couples the dynamics of the slow moving chains.

A second intermolecular intermonomer potential is the direct consequence of the chains being entangled. This potential limits the relative motion of two monomers belonging to two different chains, which are initially in contact and then interdiffuse freely until they experience the constraint in their dynamics due to entanglements. Both potentials depend in magnitude on the distances between the interacting units, which is evolving in time as the chains interdiffuse. In this way the system is considered in equilibrium only locally in the time domain, and the acting forces are solved self-consistently at any given time interval. Finally all the molecules that are initially correlated become uncorrelated and at long enough time intervals the sampled dynamics is Brownian.

The number of interpenetrating chains grows with chain length as $N^{1/2}$ for unentangled chains. The entangled chain dynamics is characterized by the interplay between cooperative dynamics and the presence of entanglements. Cooperative motion occurs for polymers comprised in the volume defined by the distance between two entanglements, $d$. At length scales larger than the characteristic entanglement distance, $d$, entanglements suppress many-chain cooperative motion. The number of interacting chains  scales as $n\propto \sqrt{N_e}$ in entangled samples, and is constant for the weakly entangled samples we study. 

 It is reasonable to hypothesize that for chains longer than the ones described in this study, a secondary correlated motion could occur on the longer lengthscale of the chain size where the cooperativity would involve the domains on the entanglement lengthscale and larger in a hierarchical clustering of cooperative motion. Still, we expect the dynamics of any entangled polymer chain to exhibit cooperative motion, with a constant $n\propto \sqrt{N_e}$, in the short-time regime and at length scales shorter than $d$.

It is noteworthy that while both Rouse and reptation models focus on the dynamics of a single chain in the field of the surrounding polymers, the role played by interchain cooperativity has emerged as a key component of polymer dynamics in several recently-published experimental studies.\cite{Richter2024,Monkenbusch2023,Sharma2022a,Richter2021a} These studies investigated both unentangled and entangled polymer dynamics, as well as dynamics of short tracer chains in an entangled matrix. They showed that chain cooperativity is present in polymer melts even when the liquid is formed by short chains: a finding that agrees with our model and our simulations.\cite{Guenza1999, Guenza2022} Interpenetration and effective chain interactions have been found to modify the dynamics of intrinsically disordered proteins in coacervates, as cooperative effects are observed in the correlated subdiffusive dynamics of their centers-of-mass.\cite{Galvanetto2023,Guenza2023}

\section{Supporting Information}
The Supporting Information contains the background derivation of the CDGLE approach for entangled dynamics; the CDGLE's solution by a similarity transformation; the time correlation functions for the entangled dynamics.

\section{Acknowledgements}
This material is based upon work supported by the National Science Foundation under Grant No. CHE-2154999.
The computational work was partially performed on the
supercomputer Expanse at the San Diego Supercomputer Center, with the support of ACCESS\cite{access} allocation Discover ACCESS CHE100082 (ACCESS is a program supported by the National Science Foundation under Grant No. ACI-1548562). This work also benefited from access to the University of Oregon high performance computing cluster, Talapas.
M.G.G. thanks James Donley for the careful reading of the manuscript.

\clearpage

%\bibliography{Ref}
\bibliography{EsubmissionNew}
\end{document}